\documentclass[journal,comsoc]{IEEEtran}

\usepackage[T1]{fontenc}

\usepackage{times}
\usepackage{amsmath}  
\usepackage{amssymb}  
\usepackage{mathrsfs} 
\usepackage{theorem}  
\usepackage{cite}     
\usepackage{comment}  

\usepackage{upref}
\usepackage{amsfonts}

\usepackage{verbatim}

\usepackage[dvipsnames,usenames]{color}

\usepackage{graphicx}
\usepackage{subfigure}
\usepackage{psfrag}
\usepackage{float}
\restylefloat{table}
\usepackage{multirow}
\usepackage{epsfig}
\usepackage{algorithmicx}
\usepackage{algorithm}
\usepackage{algpascal}
\usepackage{algcompatible}
\usepackage{algpseudocode}

\newcommand{\be}[1]{\begin{equation}\label{#1}}
\newcommand{\ee}{\end{equation}}

\newcommand{\bc}{\begin{center}}
\newcommand{\ec}{\end{center}}

\newcommand{\cA}{{\cal A}}
\newcommand{\cB}{{\cal B}}

\newcommand{\cL}{{\cal L}}

\newcommand{\cN}{{\cal N}}

\newcommand{\bfe}{{\boldsymbol e}}

\newcommand{\bfv}{{\boldsymbol v}}

\newcommand{\bfx}{{\boldsymbol x}}
\newcommand{\bfy}{{\boldsymbol y}}

\newcommand{\bfR}{{\mathbf R}}

\newcommand{\bfX}{{\mathbf X}}

\renewcommand{\leq}{\leqslant}

\renewcommand{\geq}{\geqslant}

\newcommand{\prob}{\text{Pr}}

\newcommand{\Cref}[1]{Co\-rol\-la\-ry\,\ref{#1}}

\newcommand{\Figref}[1]{Fig. \ref{#1}}

\newcommand{\Tabref}[1]{Table \ref{#1}}
\newcommand{\Secref}[1]{Section \ref{#1}}
\newcommand{\Equref}[1]{equation (\ref{#1})}

\theoremstyle{plain} \theorembodyfont{\normalfont\slshape}

\newtheorem{thm}{Theorem$\!$}

\newtheorem{prop}{Proposition$\!$}

\newtheorem{lem}{Lemma$\!$}

\newtheorem{conj}[thm]{Conjecture$\!$}

\newtheorem{cor}{Corollary$\!$}

\newtheorem{cl}{Claim$\!$}

\newtheorem{defi}{Definition$\!$}

\newtheorem{const}{Construction$\!$}

\theorembodyfont{\normalfont}

\newtheorem{exam}{Example$\!$}
\newenvironment{example}{\begin{exam}\hspace*{-1ex}{\bf .}}{\end{exam}}

\newtheorem{remrk}{Remark$\!$}

\newlength{\paragraphindent}
\newlength{\widthone}
\newlength{\widthtwo}
\newlength{\widththree}
\newlength{\colwidthtemp}
\definecolor{Codecolor}{named}{White}  

\newcommand{\Copen}{\mbox{\{\kern-5.50pt\{}}
\newcommand{\Cclose}{\mbox{\}\kern-5.50pt\}}}
\newcommand{\Cslash}{\mbox{$\backslash\kern-6.02pt\backslash$}}

\begin{document}
%
\title{Multihead Multitrack Detection with ITI Estimation in Next Generation Magnetic Recording System}

\author{Bing Fan,~\IEEEmembership{Student Member,~IEEE,}
        Hemant K. Thapar,~\IEEEmembership{Fellow,~IEEE,}
        and Paul H. Siegel,~\IEEEmembership{Fellow,~IEEE}
\thanks{B. Fan and P. H. Siegel are with the Department of Electrical \& Computer
Engineering, University of California, San Diego, CA 92093 USA (e-mail:
bifan@ucsd.edu; psiegel@ucsd.edu)}
\thanks{H. Thapar is with OmniTier Storage (e-mail: hemantkthapar@gmail.com)}}


\maketitle

\begin{abstract}
 Multitrack detection with array-head reading is a promising technique proposed for next generation magnetic storage systems. The multihead multitrack (MHMT) system is characterized by intersymbol interference (ISI) in the downtrack direction and intertrack interference (ITI) in the crosstrack direction. Constructing the trellis of a MHMT maximum likelihood (ML) detector requires knowledge of the ITI, which is generally unknown at the receiver. In addition, to retain efficiency, the ML detector requires a static estimate of the ITI, whose true value may in reality vary. In this paper we propose a modified ML detector on the $n$-head, $n$-track ($n$H$n$T) channel which could efficiently track the change of ITI, and adapt to new estimates. The trellis used in the proposed detector is shown to be independent of the ITI level. A gain loop structure is used to estimate the ITI. Simulation results show that the proposed detector offers a performance advantage in settings where complexity constraints limit the traditional ML detector to use a static ITI estimate. 
 
\end{abstract}
\IEEEpeerreviewmaketitle
\section{Introduction}
\label{sec_intro}
With the development of information networks and data centers, the demand for ultra-high capacity storage devices is continuously increasing. In the next generation magnetic recording systems, data tracks are squeezed to be closer and thinner, to achieve higher areal density. In the readback process, the read head can sense signals from adjacent tracks when reading from the target track, causing intertrack interference (ITI)\cite{Wood_TDMR}\cite{NNM05}. This additional noise source could heavily degrade the performance of disk drives using conventional detection methods\cite{Roh_singlehead}.

Several techniques, based on different practical requirements, have been proposed to resolve the ITI problem. The performance of two single-head/single-track (SHST) detectors are studied in \cite{Roh_singlehead}. Iterative ITI cancellation, which removes ITI from each single-track readback signal before detection, is explored in \cite{kumar_ITIcancelation} and \cite{Fujii_multitrack}. These SHST techniques maintain acceptable performance when ITI is low, but suffer as ITI becomes severe. Multihead multitrack (MHMT) schemes have attracted considerable attention because of their ability to better combat ITI. It can be achieved by using an array of heads to read multiple tracks simultaneously, or by using one head to sequentially scan a group of tracks. These readback signals are processed together to make a decision on the target tracks. The advantage of using MHMT format is tested and theoretically analyzed in \cite{barbosa_simultaneously} and \cite{Soljanin_multihead}. An iterative detection/decoding scheme for a two-track channel model with two heads is simulated in \cite{Ma_iterative}. In \cite{ZVKZ14}, the authors study the performance and implementation cost of MHMT detector for shingled magnetic recording (SMR), and similar structures are also analyzed for bit patterned media (BMP) recording in \cite{KSW10} and \cite{KSWB10}.

Our work is first developed on a linear and symmetric two-head/two-track (2H2T) model such as that used in \cite{Fujii_multitrack} \cite{barbosa_simultaneously}\cite{Soljanin_multihead}\cite{Ma_iterative}. One problem associated with ITI is how to estimate the response from an adjacent track. The authors of \cite{Roh_singlehead} propose a least mean square (LMS) adaptive algorithm to estimate the off-track interference for the SHST system. For the 2H2T case, we reformulate this parameter estimation problem as a gain control model, and propose a novel detector -- the weighted sum-subtract joint detector (WSSJD) -- along with a gain loop to adaptively estimate the ITI level. The proposed algorithm keeps the ML merit, and relaxes the constraints of using a static ITI estimate in the traditional ML detector. Part of this work was introduced in \cite{bing_ICC}.

Another important issue associated with an optimal  maximum-likelihood (ML) MHMT detector is its high computational complexity,  which is proportional to $ 2^{M\nu} $, where $ M $ is the number of tracks jointly processed, and $ \nu $ is the channel memory. For a system that jointly detects many tracks  or that has a long channel impulse response, an ML detector will be impractical. The WSSJD technique offers a natural set partition principle in the input constellation, and a reduced complexity implementation can be applied. This concept is presented in \cite{bing_intermag}.

The proposed algorithm can be generalized to $n$-head, $n$-track model by taking the eigenvalue decomposition of the interference matrix, and applying coordinate transformation both in the input space and the output space. After the decomposition, the ITI channels are transformed into $n$ separate and parallel channels. The ITI level appears as a gain factor of each resulting channel, and can be estimated by the gain loops. We present the simulation results for a 3H3T system.

The paper is organized as follows. \Secref{sec_twohead} introduces the 2H2T system model and reviews the optimal detector. In \Secref{sec_ssjd} we present the WSSJD and analyze its performance in terms of a minimum distance parameter. In \Secref{sec_gainloop} we study the ITI sensitivity of the detectors. We also analyze the effect of ITI mismatch on different types of error events. A gain loop structure is then proposed to adaptively estimate the ITI level for the use by WSSJD. We show the performance of proposed algorithm in \Secref{sec_simulation}, where we consider both static and adaptive ITI environments. We also present the performance of a reduced-complexity implementation of WSSJD on the EPR4 channel. We generalize WSSJD to $n$-head, $n$-track system in \Secref{sec_general}, and conclude the paper in \Secref{sec_conclusion}.


\section{Two Head/Two Track System}
\label{sec_twohead}
\begin{figure}[!t]
\centering
\includegraphics[width=.6\columnwidth]{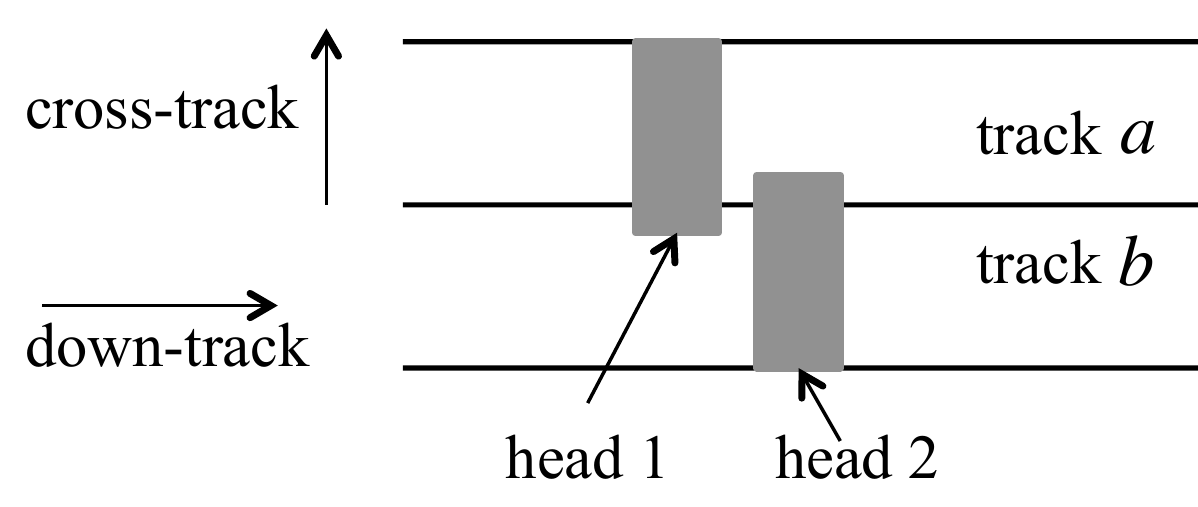}  
\caption{Schematic of a two-head/two-track recording system}
\label{fig_twohead}
\end{figure}
We consider a linear and symmetric 2H2T system as shown in \Figref{fig_twohead}. Track $ a $ and track $ b $ are two adjacent tracks with no guard band between them. Let $ x^a(D)$, $x^b(D)$ be the data sequences recorded on tracks $a$ and $b$, with $x^i(D)=\Sigma_{k=0}^N\,x_k^iD^k$ and $x_k^i\in\{-1,+1\}$ for $i\in\{a,b\}$. We assume $ x^a(D)$, $x^b(D)$ are both i.i.d. and equiprobable, and $x^b(D)$ is independent of $x^a(D)$. We also assume that there is no phase offset during the writing, i.e., the written patterns $ x^a(D)$, $x^b(D)$ are perfectly aligned. Head 1 and head 2 have the same dimensions, are placed symmetrically over track $ a $ and track $ b $,  and  move together in the down-track direction. 

During readback, the signal from each head is passed through a matched filter, a sampler, and then equalized to the target dipulse response represented by polynomial $h(D)=h_0+h_1D+\dots+h_{\nu}D^{\nu}$ of degree $\nu$. The interference from the side track is additive and formulated as a scaled output from the ISI channel $h(D)$. The noiseless outputs of the 2H2T channel are given by
\begin{align}
y^a(D)=x^a(D)h(D)+\epsilon\, x^b(D)h(D) \notag
\\ y^b(D)=\epsilon\, x^a(D)h(D)+x^b(D)h(D) \label{eq_y_output}
\end{align}
where $\epsilon$ represents the ITI level determined by the overlap between the head and the side track. 

The received signals from head 1 and head 2 are further corrupted by the electronic noise, i.e.
\begin{align}
r^a(D)=y^a(D)+n^a(D) \notag
\\ r^b(D)=y^b(D)+n^b(D) \label{eq_noisy_output}
\end{align}
where $n^{a}(D),n^{b}(D)$ are uncorrelated and i.i.d sequences, with $n_{k}^{a},n_{k}^{b}\sim\cN(0,\sigma^{2})$.

At the receiver, the ML detector makes a decision by finding the input pair $\hat{x}^a(D)$, $\hat{x}^b(D)$ that maximize the log likelihood of the received signals, i.e.
\begin{align}
&\hat{x}^a(D),\hat{x}^b(D) \notag \\ 
&=\underset{x^a,x^b}{\arg\max}\ \log \prob(r^a(D),r^b(D)\,|\,x^a(D),x^b(D)) \notag \\
					  &=\underset{x^a,x^b}{\arg\min}\ \|r^a(D)-y^a(D)\|^2 +\|r^b(D)-y^b(D) \|^{2}
\end{align}
where $\|\cdot\|^2$ denotes the squared Euclidean norm,
\[
\|x(D)\|^2=\sum_k\,x_k^2.
\] 
In other words, the received sequences are jointly decoded to the sequence pair whose noiseless channel outputs are closest to the received signals in the output space. This can be done by passing the received signals through a two-track Viterbi detector. The trellis is designed to simultaneously recover both tracks. Each trellis edge goes from an initial state $ s(k-1)= [x^a_{k-\nu}\dots x^a_{k-1},\,x^b_{k-\nu}\dots x^b_{k-1}] $
 to a terminal state $ s(k)= [x^a_{k-\nu+1}\dots x^a_{k},\,x^b_{k-\nu+1}\dots x^b_{k}] $
with input label $\cL_{\text{in}}=(x^a_{k},x^b_{k})$ and output label $\cL_{\text{out}}=(y^a_{k},y^b_{k})$. 
 For a channel with memory $\nu$, the trellis contains $2^{2\nu}$ states each of which is associated with $4$ incoming and outgoing edges. 
 
The ML detector needs to know the value  $\epsilon$  to calculate the noiseless output label $(y^a_{k},y^b_{k})$ given by \Equref{eq_y_output}. Therefore, the conventional ML detector works efficiently only when $\epsilon$ is static. For varying $\epsilon$, the conventional ML detector has to recalculate the output label $(y^a_{k},y^b_{k})$ whenever the value of $\epsilon$ changes. If the channel trellis has a large number of branches or if $\epsilon$ changes continuously, this adaptation process incurs considerable delay. On a real hard drive, however, $\epsilon$ generally varies spatially due to mechanical effects such as head skew and flying height variation. Thus, adaptive estimation of $\epsilon$ will be necessary, introducing significant detection latency.
 
In the following sections, we present a novel detection architecture that makes it possible to adaptively estimate $\epsilon$ while retaining the efficiency of ML detection. We show that the proposed approach achieves ML performance with static ITI, but has the flexibility to efficiently work with an adaptive estimator for the ITI level $\epsilon$. The proposed detector uses a different trellis diagram than the conventional two-track ML detector.  For convenience, we refer to the latter as the ``ML trellis'' even though both detectors produce the ML output sequences.

Let $[x^a(D), x^b(D)]$ and $[\hat{x}^a(D), \hat{x}^b(D)]$ be the correct and estimated sequences, respectively. An error event happens if $e^a(D) = x^a(D)-\hat{x}^a(D)$ and $e^b(D) = x^b(D)-\hat{x}^b(D)$ are not zero. The distance parameter of a given error event is calculated by 
\begin{align}
d^2(e^a(D), e^b(D)) = &\|e^a(D)h(D)+\epsilon\, e^b(D)h(D)\|^2 \notag \\&+ \|\epsilon \, e^a(D)h(D)+ e^b(D)h(D)\|^2
\end{align}
It is well known that the error event probability of the trellis-based detector can be approximated as $\text{Pe} \propto Q(\frac{d_{\text{min}}}{2\sigma})$,
where  the  Q-function is the tail probability of the standard Gaussian distribution, $d_{\text{min}}$ is the minimum distance parameter over all possible error events, and $\sigma$ is the standard deviation of the additive Gaussian channel noise.  The performance of the detector can be accurately predicted by analyzing the minimum distance. As given in \cite{Soljanin_multihead}, the minimum distance parameter of the ML detector on the 2H2T channel is 
 \begin{align}
  d_{\text{min, ML}}^2 = \left\{ 
   \begin{array}{l l}
     (1+\epsilon^2)d_0^2 & \quad \text{if $0\leq \epsilon \leq 2-\sqrt{3}$}\\
     2(1-\epsilon)^2d_0^2 & \quad \text{if $2-\sqrt3\leq \epsilon \leq 1/2$}
   \end{array} \right.
   \label{eq_3}
 \end{align}
 where $d_0$ is the minimum distance of a single track with channel polynomial $h(D)$ when there is no ITI. When ITI is low, the single track error events are the minimum distance error patterns. When ITI increases, the double track error events become the dominant error events. The operating point that gives the highest minimum distance, or the best performance of the ML detector, is at $\epsilon=2-\sqrt{3}$. 
 

\section{Weighted Sum-Subtract Joint Detection}
\label{sec_ssjd}
The weighted sum-subtract joint detection (WSSJD) algorithm differs from the conventional ML detector in two respects. First, it adds a ``sum-subtract'' preprocessor before the Viterbi detector. Second, it uses weighted branch metrics in the Viterbi detector. When we introduce the algorithm, we assume $\epsilon$ to be known. This condition will be relaxed in \Secref{sec_gainloop} where we show that $\epsilon$ acts as a gain factor that can be estimated by means of a first-order gain loop.
\subsection{Sum-subtract preprocessing}
Instead of directly passing the received sequences $r^a(D)$ and $r^b(D)$ to the Viterbi detector, the WSSJD first calculates their sum $r^+(D)$ and difference $r^-(D)$, normalized by $\frac{1}{1+\epsilon}$ and $\frac{1}{1-\epsilon}$, respectively, i.e.,  
\begin{align}
r^+(D)=\frac{1}{1+\epsilon}\,(r^a(D)+r^b(D)) \notag
\\ r^-(D)=\frac{1}{1-\epsilon}\,(r^a(D)-r^b(D)). \label{eq_ss}
\end{align}
Defining the sum and difference input signals by
\begin{align}
z^+(D)=x^a(D)+x^b(D), \, \,  z^-(D)=x^a(D)-x^b(D),\label{eq_ss_1}
\end{align}
and the corresponding noiseless output signals by 
\begin{align}
y^+(D)=z^+(D)h(D),  \, \,  y^-(D)=z^-(D)h(D). \label{eq_ss_3}
\end{align}
We can rewrite \Equref{eq_ss} as
\begin{align}
r^+(D)=y^+(D)+n^+(D) \notag
\\ r^-(D)=y^-(D)+n^-(D)\label{eq_ss_2}
\end{align}
where the Gaussian noise components 
\begin{align}
n^+(D)=\frac{1}{1+\epsilon}\,(n^a(D)+n^b(D)), \notag\\
n^-(D)=\frac{1}{1-\epsilon}\,(n^a(D)-n^b(D))\,\label{eq_noise}
\end{align}
satisfy $n^+_k\sim\cN(0, \frac{2\sigma^2}{(1+\epsilon)^2})$, $n^-_k\sim\cN(0, \frac{2\sigma^2}{(1-\epsilon)^2})$. 
Furthermore, 
\begin{align}
E(n^+_k\,n^-_k)=\frac{1}{1-\epsilon^2}(E({n^a_k}^2)-E({n^b_k}^2))=0
\end{align}
which implies that $n^+(D)$ and $n^-(D)$ are uncorrelated and, therefore,  independent.

We can think of $r^+(D)$ and $r^-(D)$ as the noisy outputs obtained by  passing each of $z^+(D)$ and $z^-(D)$ through a channel $h(D)$, but with different SNRs. These channels are called the ``sum channel'' and the ``subtract channel,'' respectively. Notice that the corresponding input  sequences $z^+(D)$ and $z^-(D)$ have a three-level alphabet, $\cB=\{-2,0,2\}$. There is a one-to-one mapping between $(z^+_k,z^-_k)$ and $(x^a_k,x^b_k)$, as shown in \Tabref{table_1}.

\begin{table}
  \caption{mapping between $(x^a_k,x^b_k)$ and $(z^+_k,z^-_k)$}
  \label{table_1}
\centering
  \begin{tabular}{ r r || r r }
    \hline
    $x^a_k$ & $x^b_k$ & $z^+_k$ & $z^-_k$  \\ \hline
    1 & 1 & 2 & 0 \\ 
    1 & -1 & 0 & 2 \\ 
   -1 & 1 & 0 & -2 \\
   -1 & -1 &  -2 & 0  \\  
    \hline
  \end{tabular}
\end{table}

Since $r^+(D)$ and $r^-(D)$ are obtained from separate channels, one can independently detect $z^+(D)$ and $z^-(D)$, and then map $(z^+_k,z^-_k)$ to $(x^a_k,x^b_k)$ according to \Tabref{table_1}. This method corresponds to solving two detection problems
\begin{align}
\hat{z}^+(D)=\underset{z^+}{\arg\max}\log \prob(r^+(D)\,|\,z^+(D)) \notag \\
             =\underset{z^+}{\arg\min}\| r^+(D)\, - \,z^+(D)\|^2\, \notag \\
\hat{z}^-(D)=\underset{z^-}{\arg\max}\log \prob(r^-(D)\,|\,z^-(D)) \notag\\
=\underset{z^+}{\arg\min}\| r^-(D)\, - \,z^-(D)\|^2. 
\end{align}

However, this approach is not optimal. From \Tabref{table_1} we see that $z^+(D)$ and $z^-(D)$ are not independent, e.g., $z^+_k=2$ forces $z^-_k$ to be $0$. Independent detection ignores this correlation and produces some undecodable $(\hat{z}^+_k, \hat{z}^-_k)$ pairs. Optimal detection must jointly consider both the sum channel and the subtract channel, determining

\begin{align}
&\hat{z}^+(D),\hat{z}^-(D) \notag \\
& =\underset{z^+,z^-}{\arg\max}\log \prob(r^+(D),r^-(D)\,|\,z^+(D),z^-(D)). \label{max_prob}
\end{align}

\begin{figure}
 \centering
 \includegraphics[width=.75\columnwidth]{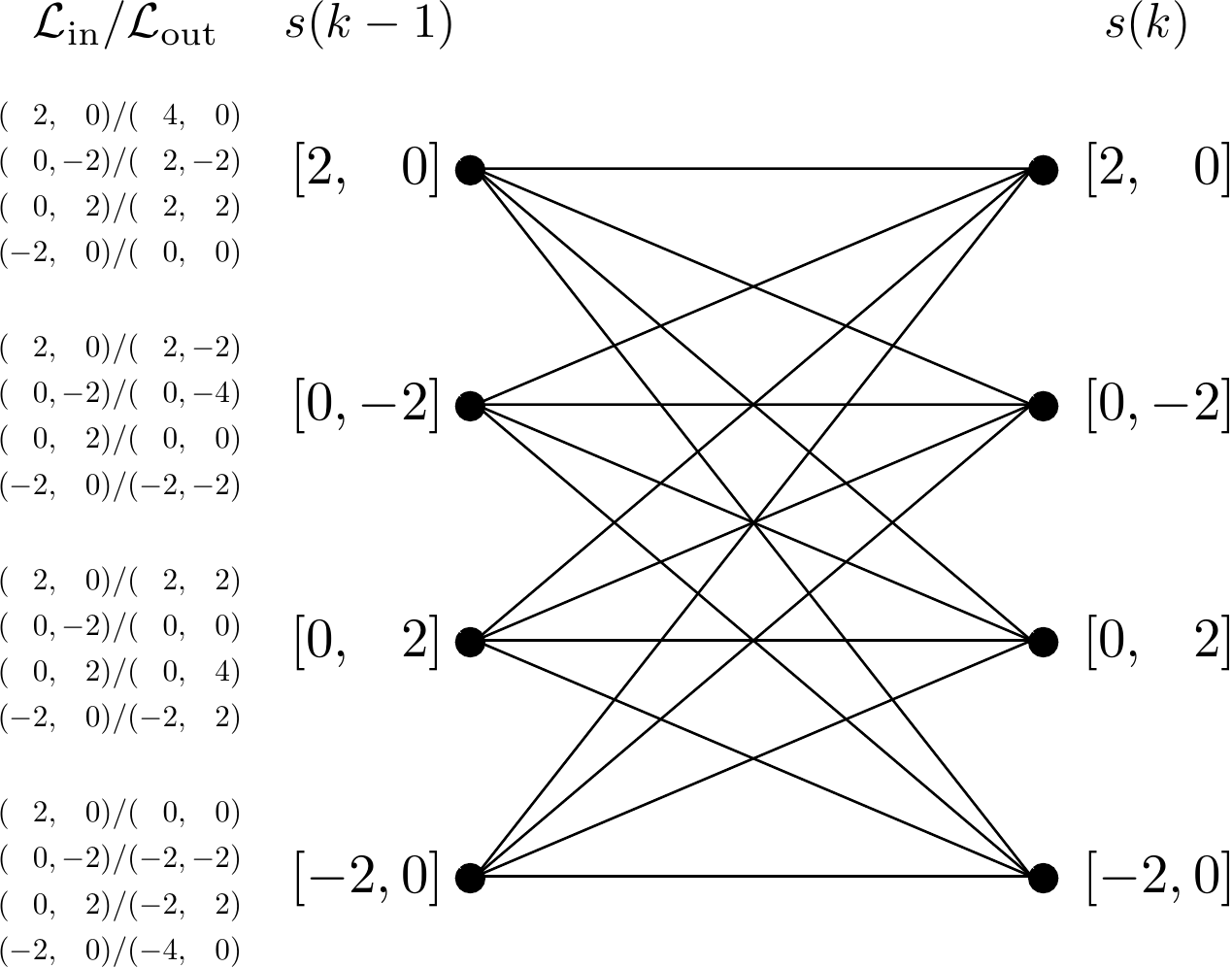}
 \caption{WSSJD trellis for channel $h(D)=1+D$}
 \label{fig_ssjdtrellis}
\end{figure}

The WSSJD provides a practical trellis-based algorithm for solving this problem.
The WSSJD trellis has the same number of states as the ML trellis. Each branch connects an initial state $s(k-1)=[z^+_{k-\nu}\dots z^+_{k-1}, z^-_{k-\nu}\dots z^-_{k-1}]$ to a terminal state $s(k)=[z^+_{k-\nu+1}\dots z^+_{k}, z^-_{k-\nu+1}\dots z^-_{k}]$ with input label $\cL_{\text{in}}=(z^+_k,z^-_k)$ and output label $\cL_{\text{out}}=(y^+_k,y^-_k)$. \Figref{fig_ssjdtrellis} shows a WSSJD trellis for the channel $h(D)=1+D$. The text to the left of each state lists the branch labels in the form of input/output. Note that, unlike the ML trellis, the WSSJD trellis is independent of $\epsilon$.
 
\subsection{Weighted branch metric}
Since the sum channel and the subtract channel have different noise powers, the WSSJD computes a weighted sum of their individual distance metrics, $\|r^+(D)-y^+(D)\|^2$ and $\|r^-(D)-y^-(D)\|^2$.
The optimal choice of the weights is found by evaluating \Equref{max_prob}:
\begin{align}
& \hat{z}^+(D),\hat{z}^-(D) \notag\\ 
&=\underset{z^+,z^-}{\arg \max}\,\log \prob(r^+(D),r^-(D)|z^+(D),z^-(D)) \notag\\
&=\underset{z^+,z^-}{\arg \max}\,\log \prob(r^+(D)|z^+(D))+\log \prob(r^-(D)|z^-(D)) \notag\\
&=\underset{z^+,z^-}{\arg \min}\,\frac{\| r^+(D)-y^+(D)\|^2}{2\sigma^2/(1+\epsilon)^2}+\frac{\| r^-(D)-y^-(D)\|^2}{2\sigma^2/(1-\epsilon)^2} \notag \\
&=\underset{z^+,z^-}{\arg \min}\,(1+\epsilon)^2\|  r^+(D)-y^+(D)\|^2\notag\\
&\qquad \qquad \quad+(1-\epsilon)^2 \| r^-(D)-y^-(D)\|^2. \label{eq_weights}
\end{align}
Let $M_{k-1}(s)$ denote the survivor path metric for state $s$ at time $k-1$. Then \Equref{eq_weights} suggests that the path metric corresponding to the extension along a branch from state $s$ to state $s'$ at time $k$ is 
\begin{multline}
M_{k}(s')=M_{k-1}(s)+(1+\epsilon)^2(r_k^+-y_k^+)^2 \\ +(1-\epsilon)^2(r_k^--y_k^-)^2
\label{eq_extended_path_metric}
\end{multline}
where $(y_k^+, y_k^-)$ is the output label of the branch. The term $m(s, s')=(1+\epsilon)^2(r_k^+-y_k^+)^2+(1-\epsilon)^2(r_k^--y_k^-)^2$ is called the weighted branch metric.

Since the transformation in the sum-subtract preprocessing is bijective, we have 
\begin{multline}
\prob(r^+(D),r^-(D)|z^+(D),z^-(D)) \\ =\prob(r^a(D),r^b(D)|x^a(D),x^b(D)). \label{eq_MLequ}
\end{multline}
Therefore WSSJD gives the ML solution.

Assume $(z^+(D),z^-(D))$ are the correct input sequences. WSSJD outputs wrong estimates $(\hat{z}^+(D),\hat{z}^-(D))$ if
\begin{multline}
\ \prob(r^+(D),r^-(D)|z^+(D),z^-(D))\\<\prob(r^+(D),r^-(D)|\hat{z}^+(D),\hat{z}^-(D)), \label{eq_16}
\end{multline}
Let $e^+(D)=z^+(D)-\hat{z}^+(D)$ and $e^-(D)=z^+(D)-\hat{z}^-(D)$ be the error event. Notice that the alphabet for $e^+_k$ and $e^-_k$ is $\{\pm 4, \pm 2,0 \}$, and $e^+_k$ and $e^-_k$ are not independent, e.g. $e^+_k = 4$ implies $e^- = 0$. The probability of having $(e^+(D),e^-(D))$, given $z^+(D)$ and $z^-(D)$ are the recorded sequences, equals to $Q(\frac{d_{\text{WSSJD}}(e^+(D),e^-(D))}{2\sigma})$, where
\begin{align}
& d^2_{\text{WSSJD}}(e^+(D),e^-(D)) \notag \\&=\frac{(1+\epsilon)^2\|e^+(D)h(D)\|^2+(1-\epsilon)^2\|e^-(D)h(D)\|^2}{2} \label{eq_17}
\end{align}
is the effective distance parameter defined for WSSJD. Although the WSSJD trellis is independent of $\epsilon$, its distance measure is redefined by considering the effect of SNR differences in the sum and subtract channels, in order to make a fair comparison with other detectors. Evaluating equation (\ref{eq_17}) for all possible error events shows that WSSJD has the same minimum distance parameter as the ML detector.
\begin{figure}
\centering
\includegraphics[width=.9\columnwidth]{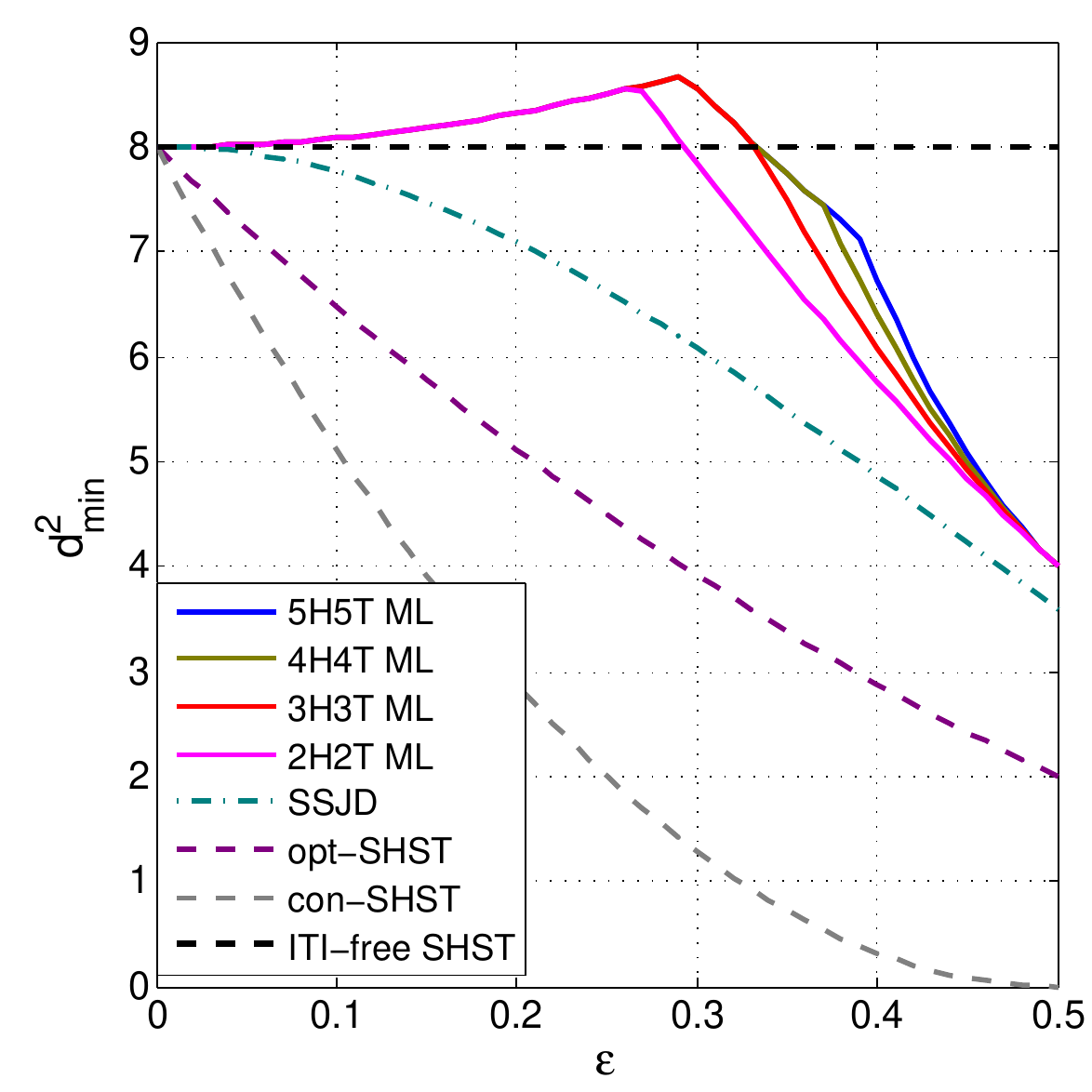}
\caption{Minimum squared distance comparison of different detectors on channel $h(D)=1+D$ with $d_0^2=8$. }
 \label{fig_distance}
\end{figure}
\subsection{Performance loss of unweighted branch metric}
The detector that ignores the weighting factors, i.e., that uses 
\begin{align}
m(s, s')=(r_k^+-y_k^+)^2+(r_k^--y_k^-)^2
\end{align}
 as the  branch metric, is suboptimal. We refer to this as sum-subtract joint detection (SSJD). The performance loss incurred by SSJD is reflected in its minimum distance parameter. Let $(z^+(D), z^-(D))$ and $(\hat{z}^+(D), \hat{z}^-(D))$ be the correct sequences and estimated sequences. The error event probability is
\begin{align}
& \text{Pr}(\|r^+(D)-z^+(D)h(D)\|^2 + \|r^-(D)-z^-(D)h(D)\|^2 \notag\\ & \quad > \|r^+(D)-\hat{z}^+(D)h(D)\|^2 + \|r^-(D)-\hat{z}^-(D)h(D)\|^2 ) \notag \\
& = Q(\frac{d_{\text{SSJD}}(e^+(D), e^-(D))}{2\sigma}).
\end{align}
where
\begin{align}
d_{\text{SSJD}}(e^+(D), e^-(D))=\frac{\|e^+(D)h(D)\|^2 + \|e^-(D)h(D)\|^2}{ \sqrt{\frac{2\|e^+(D)h(D)\|^2}{(1+\epsilon)^2}+\frac{2\|e^-(D)h(D)\|^2}{(1-\epsilon)^2}}}
\end{align}
Since $e^+(D)$ and $e^-(D)$ are not independent, we express them as $e^+(D)=e^a(D)+e^b(D)$ and $e^-(D)=e^a(D)-e^b(D)$ to find $d^2_{\text{min, SSJD}}$. To simplify the notation, let $A(D)=e^a(D)h(D)$ and $B(D)=e^b(D)h(D)$. We have
\begin{align}
& d^2_{\text{SSJD}}(e^a(D), e^b(D)) \notag \\ &= \frac{(1+\epsilon)^2(1-\epsilon)^2(\|A(D)\|^2+\|B(D)\|^2)^2}{(1+\epsilon^2)(\|A(D)\|^2+\|B(D)\|^2)-4\epsilon \langle A(D), B(D)\rangle}
\end{align}
Consider the case of a single-track error event, i.e., assume $e^b(D)=0$, then 
\begin{align}
 d^2_{\text{SSJD}}(e^a(D), 0) &= \frac{(1+\epsilon)^2(1-\epsilon)^2}{(1+\epsilon^2)}\|A(D)\|^2 \notag \\ &\geq \frac{(1+\epsilon)^2(1-\epsilon)^2}{(1+\epsilon^2)}d_0^2. \label{eq_22}
\end{align}
The equality is achieved when $e^a(D)$ gives the minimum distance $d_0^2$ on channel $h(D)$.
For the case of a double-track error event, since 
\begin{align}
-\langle A(D), B(D)\rangle &\leq \|A(D)\|\,\|B(D)\| \notag \\ &\leq \frac{1}{2}( \|A(D)\|^2+\|B(D)\|^2 ),
\end{align}
we have
\begin{align}
d^2_{\text{SSJD}}(e^a(D), e^b(D)) &\geq (1-\epsilon)^2( \|A(D)\|^2+\|B(D)\|^2 ) \notag\\ &\geq 2(1-\epsilon)^2 d_0^2. \label{eq_24}
\end{align}
The equality is achieved when $e^a(D)=-e^b(D)$ and both $e^a(D)$ and $e^b(D)$ lead to minimum distance $d_0$ on ISI channel $h(D)$.
Comparison of (\ref{eq_22}) and (\ref{eq_24}) shows that, in contrast to WSSJD, the minimum distance of SSJD is always dominated by single-track error events for $\epsilon\in[0,0.5]$. Therefore
\begin{align}
d_{\text{min, SSJD}}^2=\frac{(1+\epsilon)^2(1-\epsilon)^2}{1+\epsilon^2}d_0^2.
\end{align}

In \Figref{fig_distance} we plot the squared minimum distance as a function of  $\epsilon$ for 2H2T ML and SSJD, as well as for two single track detectors\cite{soljanin_distance} included for comparison purposes. Recall that WSSJD is ML equivalent. The optimal single track detector jointly estimates both tracks based on single head outputs, and discards estimates of the data on side track. Its minimum distance is dominated by double track error events, leading to
\begin{align}
d_{\text{min, opt-SHST}}^2=(1-\epsilon)^2d_0^2.
\end{align}
The conventional single track detector treats ITI as additional electronic noise, and is thus suboptimal. For $1+D$ channel, its minimum distance is given by
\begin{align}
d_{\text{min, con-SHST}}^2=(1-2\epsilon)^2d_0^2.
\end{align}
The ITI-free SHST corresponds to the single track channel model with no ITI. It can be viewed as an upper bound of the performance of ITI cancellation detector, assuming the ITI can be perfectly removed. The distance properties of several higher order MHMT ML detectors are also plotted, which will be discussed in Section \ref{sec_general}.

The properties of WSSJD are summarized as follows. First, WSSJD is ML equivalent. Second, the WSSJD trellis  is independent of $\epsilon$,  which only affects the noise components in the independent sum and subtract channels and is taken into account by suitably weighting their respective branch metrics. This independence is the key to combining WSSJD with adaptive estimation of $\epsilon$.

\section{Adaptive ITI Level Estimation}
\label{sec_gainloop}

\subsection{ITI Sensitivity}
\label{subsec_sensitivity}
To evaluate the sensitivity of the various detectors to  a small change in the ITI level, we  introduce a small offset into our performance simulations. Suppose the nominal level is $\epsilon_0$, while the true ITI level is adjusted by an offset $\Delta\epsilon$. The new noiseless channel outputs are 
\begin{align}
y^a(D)=x^a(D)h(D)+(\epsilon_0+\Delta\epsilon)\, x^b(D)h(D)\, \notag
\\ y^b(D)=x^b(D)h(D)+(\epsilon_0+\Delta\epsilon)\, x^a(D)h(D)  \label{eq_noiseless_output}.
\end{align}
\begin{figure}
\centering
\includegraphics[width=\columnwidth]{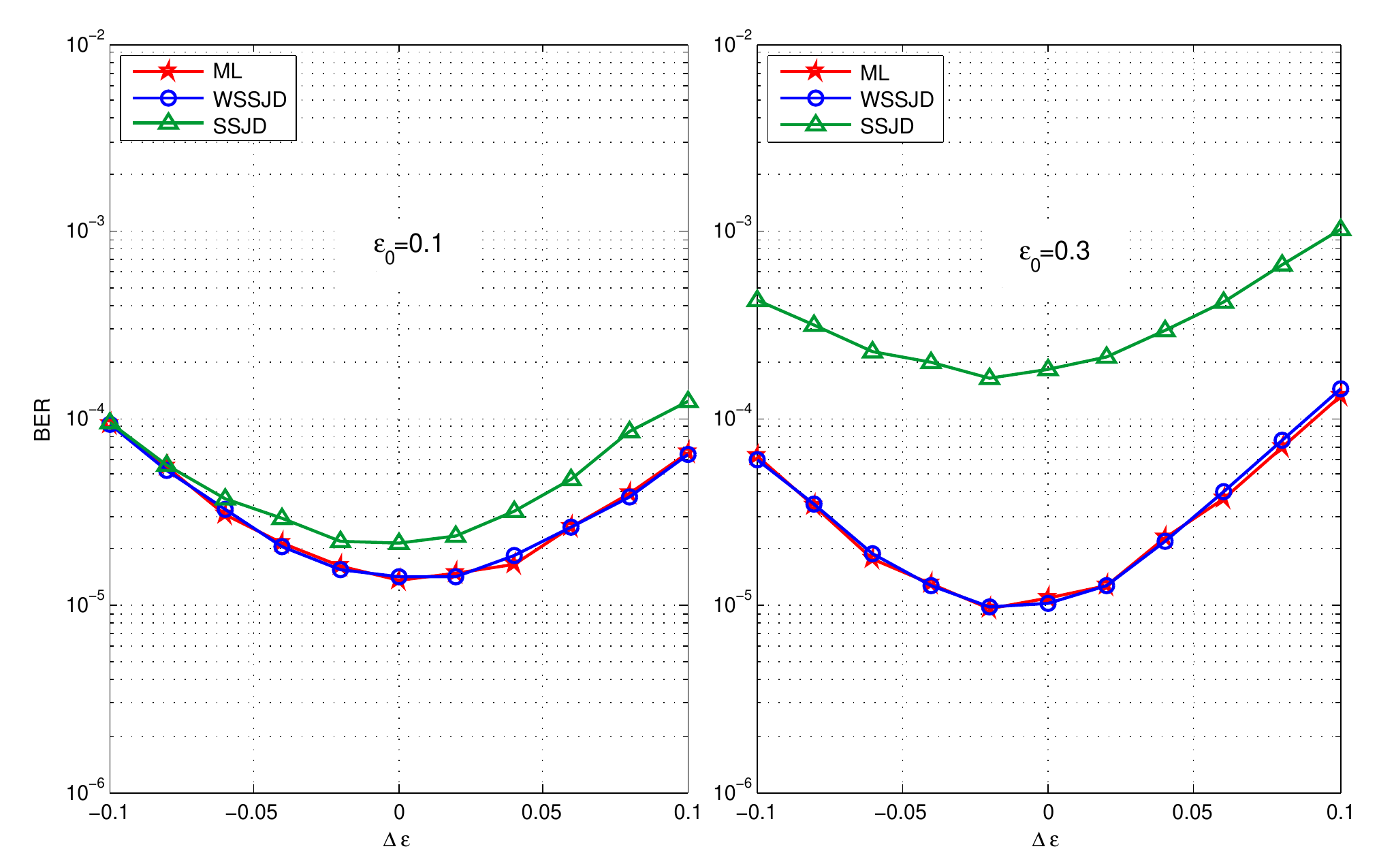} 
\caption{ITI sensitivity simulation results for different detectors with $\epsilon_0=0.1$ (left) and $\epsilon_0=0.3$ (right).}
\label{fig_asy_sim}
\end{figure}
Suppose that the detectors have inaccurate knowledge of the ITI level, and continuously use $\epsilon_0$ in detection. In this way there is a mismatch about the value of ITI level between the signal generator and the receiver. In the ideal case, $\Delta\epsilon=0$. 
\Figref{fig_asy_sim} shows the simulated bit error rate (BER) as a function of the ITI mismatch $\Delta\epsilon$ for the ML, WSSJD, and SSJD detectors on the channel $h(D)=1+D$ at  $\text{SNR}=10\text{dB}$, with $\epsilon_0 = 0.1$ and $\epsilon_0 = 0.3$, respectively. When the mismatch is small, the system performance is close to the ideal situation. We also notice that the BER curves are not symmetric about $\Delta\epsilon=0$. 
Furthermore, the minimum BER points occur at offsets with opposite polarity for $\epsilon_0=0.1$ and $\epsilon_0=0.3$. 

\Figref{fig_distance} suggests that the observed behaviors are due to minimum distance properties of the mismatched detectors.
To see this, let's take the ML detector as an example. The probability of having an error event $(e^a(D), e^b(D))$ is
\begin{align}
\text{Pe}=Q(\frac{1}{2\sigma}d(e^a,e^b,x^a,x^b))=Q(\frac{1}{2\sigma}(d_{\text{ideal}}+d_{\text{mis}})), \label{eq_30}
\end{align}
where  
{\small 
\begin{align}
& d_{\text{ideal}} =\sqrt{\| \cA(D)\|^2+\|\cB(D)\|^2}, \label{eq_31}\\ 
& d_{\text{mism}}= 2\Delta\epsilon \frac{\langle \cA(D),x^b(D)h(D)\rangle + \langle \cB(D), x^a(D)h(D) \rangle}{\sqrt{\| \cA(D)\|^2+\|\cB(D)\|^2}}, \label{eq_32}
\end{align}
\begin{align}
& \cA(D) = e^a(D)h(D)+\epsilon e^b(D)h(D), \\
& \cB(D) = e^b(D)h(D)+\epsilon e^a(D)h(D).
\end{align}}
\begin{figure}
\centering
\includegraphics[width=0.9\columnwidth]{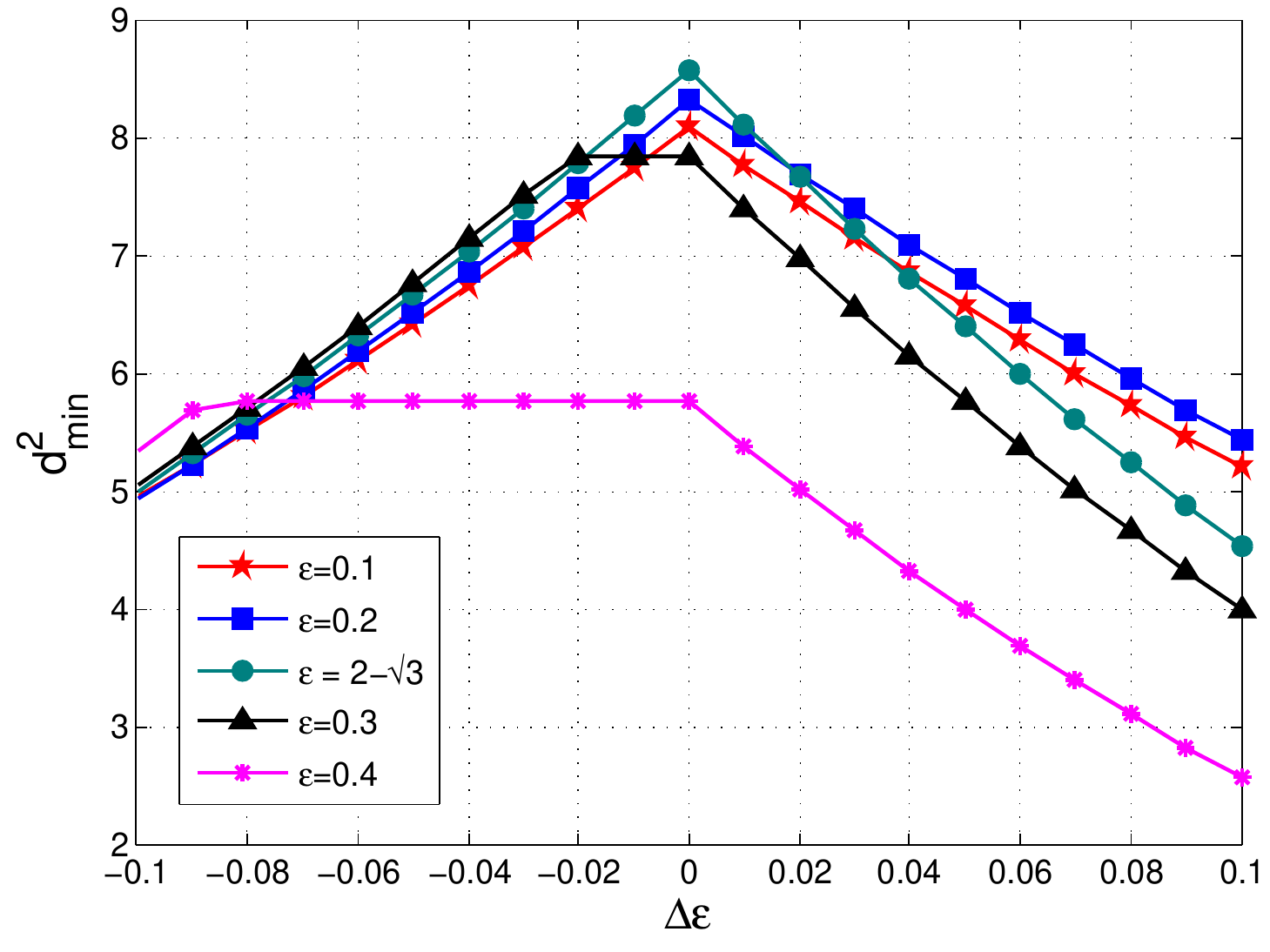}
\caption{Minimum distance parameter of the ML detector at different level of mismatch, on channel $1+D$.}
\label{fig_dismiss}
\end{figure}

Compared to the ideal case, $d_{\text{mism}}$ is the additional effect caused by the mismatch on the distance property. Notice that with the existence of mismatch, the distance parameter is now dependent of the input sequence $(x^a(D), x^b(D))$. In addition, having mismatch does not always decrease the distance. Some sequence combinations could lead to larger distance than the ideal case. The error probability is dominated by the sequence combination of $(e^a,e^b,x^a,x^b)$ that leads to the smallest value of $d_{\text{ideal}}+d_{\text{mism}}$. Finding such a combination is not an easy task because the error sequences $(e^a(D), e^b(D))$ and input sequences $(x^a(D), x^b(D))$ are independent. For example, $e^a_k=2$ forces $x^a_k$ to be $1$. Due to this correlation, it is hard to obtain an explicit expression of the minimum distance for a general channel polynomial. But for channel $1+D$, it is proved that the minimum distance of the single-track error events is
\begin{align}
 d_{\text{s}}^2 = \left\{ 
   \begin{array}{l l}
     \frac{8(1+\epsilon_0^2-2\Delta\epsilon)^2}{1+\epsilon_0^2} & \quad \text{if $\Delta\epsilon>0$}\\
     \frac{8[1+\epsilon_0^2+(2+2\epsilon_0)\Delta\epsilon]^2}{1+\epsilon_0^2} & \quad \text{if $\Delta\epsilon<0$},
   \end{array} \right.
   \label{eq_ds}
\end{align}
while with additional assistance of computer search, we show that the minimum distance of the double-track error events is
\begin{align}
 d_{\text{d}}^2 = \left\{ 
   \begin{array}{l l}
     16[(1-\epsilon_0)-2\Delta\epsilon]^2 & \quad \text{if $\Delta\epsilon>0$}\\
     16(1-\epsilon_0)^2 & \quad \text{if $\Delta\epsilon<0$}.
   \end{array} \right.
   \label{eq_dd}
\end{align}
The distance values $d^2_{\text{s}}$ and $d_{\text{d}}^2$ can be achieved by the single and double track error events that minimize $d_{\text{ideal}}$ in each case, respectively. Table \ref{table_2} and Table \ref{table_3} give examples of sequence combinations that can achieve the lower bound of $d_{\text{s}}^2$ and $d_{\text{s}}^2$ for the case of $\Delta\epsilon<0$ and $\Delta\epsilon>0$. The process to derive equation (\ref{eq_ds}) and equation (\ref{eq_dd}) is given in the appendix. The overall minimum distance of the system is 
\begin{align}
d_{\text{min}}^2=\min \, \{d_{\text{s}}^2,\,d_{\text{d}}^2\}. \label{eq_d}
\end{align}

In summary, the asymmetry of the BER curve about $\Delta\epsilon=0$ is because of the correlation between $(e^a(D), e^b(D))$ and  $(x^a(D), x^b(D))$. The reason that minimum BER points show opposite polarity at $\epsilon_0=0.1$ and $\epsilon_0=0.3$ is because at $\epsilon_0=0.1$ the system is mostly dominated by the single track error events while at $\epsilon_0=0.3$ the double track error events stand out. Fig. \ref{fig_dismiss} depicts the minimum distance found at different mismatch points for several values of $\epsilon$. Compared with Fig. \ref{fig_asy_sim}, we find that for $\epsilon_0=0.1$, a positive offset in this range tends to give higher minimum distance than a negative offset of the same magnitude. For $\epsilon_0=0.3$, this situation is reversed, and in a small range of negative offsets, $\Delta\epsilon\in[-0.02, 0]$, the mismatch doesn't reduce the minimum distance of the system. It also reduces the probability of worst case scenario, leading to a shift of the minimal BER point to the negative side.
 
\begin{table}
\centering
  \begin{tabular}{ |l | l |}
    \hline
    \multirow{3}{*}{$\Delta\epsilon<0$} & $\bfe^a=\cdots,\quad \;\; 0,\quad \! 0,\;\;\, 2,\;\;\, 0,\quad \;\; 0,\cdots$ \\
     &  $\bfx^a=\cdots,x^a_{k-2},-1,+1,-1,x^a_{k+2},\cdots$ \\
     &  $\bfx^b=\cdots,x^b_{k-2},-1,-1,-1,x^b_{k+2},\cdots$ \\
    \hline
    \multirow{3}{*}{$\Delta\epsilon>0$} & $\bfe^a=\cdots,\quad \;\; 0,\quad \! 0,\;\;\, 2,\;\;\, 0,\quad \;\; 0,\cdots$ \\
         &  $\bfx^a=\cdots,x^a_{k-2},+1,+1,+1,x^a_{k+2},\cdots$ \\
         &  $\bfx^b=\cdots,x^b_{k-2},+1,+1,+1,x^b_{k+2},\cdots$ \\
    \hline
  \end{tabular}
  \caption{Sequences achieving $d_{\text{min}}$ in equation (\ref{eq_d}) under positive/negative offset for single track error events}
\label{table_2}
\end{table}

\begin{table}
\centering
  \begin{tabular}{ |l | l |}
    \hline
    \multirow{4}{*}{$\Delta\epsilon<0$} & $\bfe^a=\cdots,\quad \;\; 0,\quad \! 0,\;\;\, 2,\;\;\, 0,\quad \;\; 0,\cdots$ \\
    & $\bfe^b=\cdots,\quad \;\; 0,\quad \! 0,-2,\;\;\, 0,\quad \;\; 0,\cdots$\\
     &  $\bfx^a=\cdots,x^a_{k-2},-1,+1,-1,x^a_{k+2},\cdots$ \\
     &  $\bfx^b=\cdots,x^b_{k-2},+1,-1,+1,x^b_{k+2},\cdots$ \\
    \hline
    \multirow{4}{*}{$\Delta\epsilon>0$} & $\bfe^a=\cdots,\quad \;\; 0,\quad \! 0,\;\;\, 2,\;\;\, 0,\quad \;\; 0,\cdots$ \\
        & $\bfe^b=\cdots,\quad \;\; 0,\quad \! 0,-2,\;\;\, 0,\quad \;\; 0,\cdots$\\
         &  $\bfx^a=\cdots,x^a_{k-2},+1,+1,+1,x^a_{k+2},\cdots$ \\
         &  $\bfx^b=\cdots,x^b_{k-2},-1,-1,-1,x^b_{k+2},\cdots$ \\
    \hline
  \end{tabular}
  \caption{Sequences achieving $d_{\text{min}}$ in equation (\ref{eq_d}) under positive/negative offset for double track error events}
\label{table_3}
\end{table}

\subsection{Gain Loop}
\begin{figure}
\centering
\includegraphics[width=0.95\columnwidth]{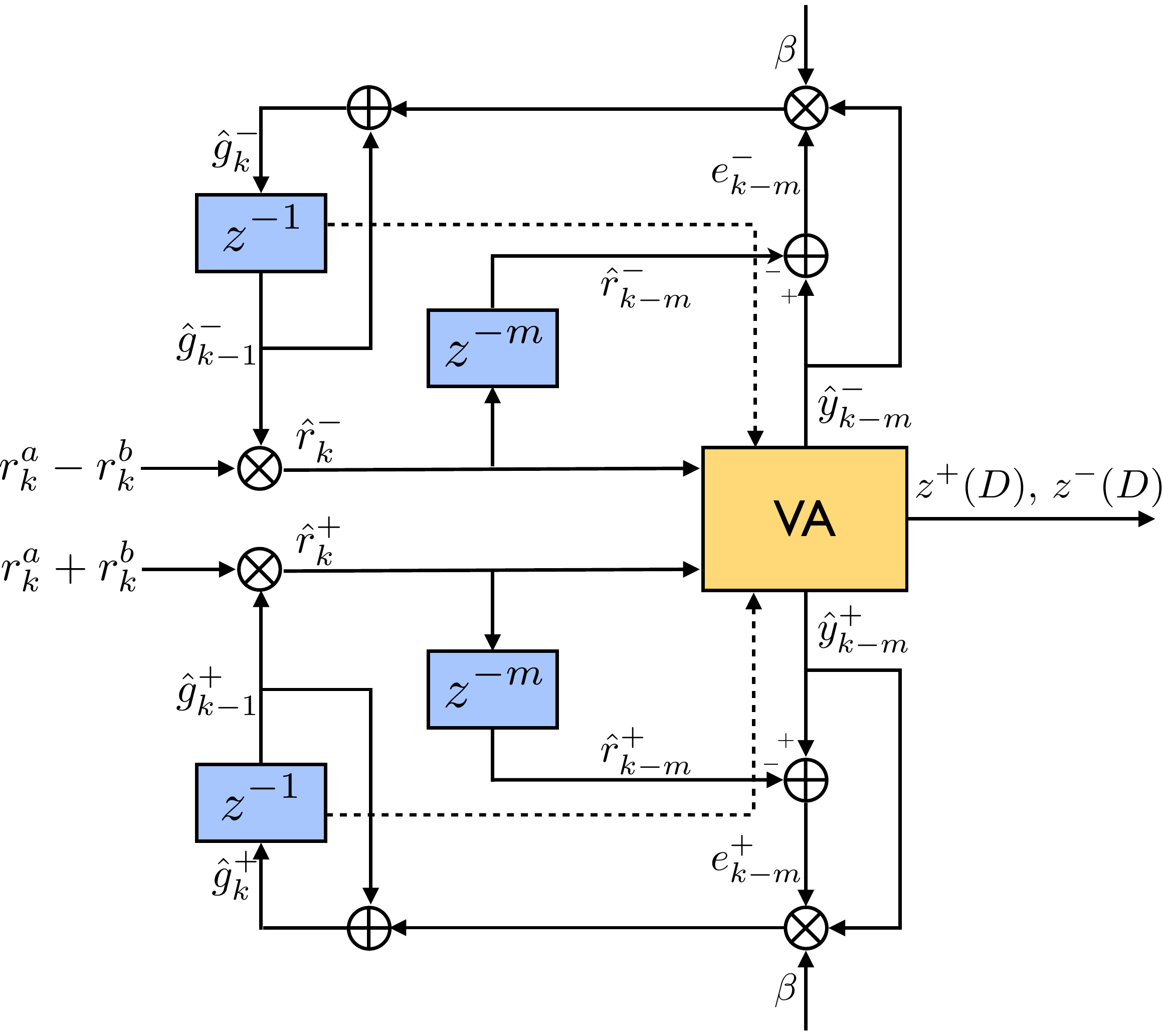} 
\caption{WSSJD with gain control to adaptively estimate ITI level }
\label{fig_gainloop}
\end{figure}
 Recall that in the sum-subtract preprocessing, $\epsilon$ appears in the gain factors that normalize signals $r^+(D)$, $r^-(D)$. We rewrite \Equref{eq_ss} as
\begin{align}
r^+(D)=g^+\,(r^a(D)+r^b(D)) \notag
\\ r^-(D)=g^-\,(r^a(D)-r^b(D)) 
\end{align}
where $g^+=\frac{1}{1+\epsilon}$, $g^-=\frac{1}{1-\epsilon}$ are the gain factors. 
%
We use the LMS adaptive algorithm to estimate these parameters. For $\hat{g}^+$, the updating rule is given by 
\begin{align}
\hat{r}^+_k &=\hat{g}^+_{k-1}\,(r^a_k+r^b_k)  \\
e_{k} & =\hat{y}^+_{k}\,-\,\hat{r}^+_{k}  \label{eq_update1}\\
\hat{g}^+_k & =\hat{g}^+_{k-1}+\beta\; \hat{y}^+_{k}\, e_{k}\label{eq_update2}
\end{align}
where $\beta$ is the step-size parameter and $\hat{y}^+_k$ is the instantaneous decision fed back from the Viterbi detector. 
The step-size parameter $\beta$ controls the convergence speed. A large $\beta$ makes the loops converge faster, but also results in larger error variance.

 One can also introduce a small delay $m\geq 1$ to get more accurate tentative decisions. In this case, \Equref{eq_update1} and \Equref{eq_update2} become
 \begin{align}
 e_{k-m} & =\hat{y}^+_{k-m}\,-\, \hat{r}^+_{k-m} \label{eq_update11}\\
 \hat{g}^+_k\quad & =\hat{g}^+_{k-1}+\beta\; \hat{y}^+_{k-m}\, e_{k-m}\label{eq_update22}
 \end{align}
The estimates $\hat{g}^+_k, \hat{g}^-_k$ will be used in the  next iteration, and also in the Viterbi detector path metric calculation~\Equref{eq_extended_path_metric}, i.e., 
 
 \begin{align}
 M_{k}(s')=M_{k-1}(s)+\hat{g}^+_{k-1}\,^2(r_k^+-y_k^+)^2 +\hat{g}^-_{k-1}\,^2(r_k^--y_k^-)^2.
 \end{align}
 
\Figref{fig_gainloop} shows a complete block diagram for WSSJD with adaptive gain estimation. The system contains two separate gain loops for $\hat{g}^+_k$ and $\hat{g}^-_k$.
While a combined loop for estimating $\hat{g}^+_k$ and $\hat{g}^-_k$ can provide a better estimate for $\epsilon$, using separate loops achieves similar performance and is more efficient.


In our simulations,  $\hat{g}_0^+$ and $\hat{g}_0^-$ are initially set to $1$. At time $k$, $r^a_k+r^b_k$ and $r^a_k-r^b_k$ are normalized by the previously estimated gain factors $\hat{g}^+_{k-1}$ and $\hat{g}^-_{k-1}$, respectively. The resulting signals $\hat{r}^+_k$ and $\hat{r}^-_k$ are sent to the Viterbi detector. The path metric of each trellis state is evaluated and scaled by $\hat{g}^+_{k-1}$ and $\hat{g}^-_{k-1}$. After comparing the path metrics, the Viterbi detector picks the most likely path, and feeds back its decision on $\hat{y}^+_{k-m}$ and $\hat{y}^-_{k-m}$. The error signal is calculated to update $\hat{g}^+_{k}$ and $\hat{g}^-_{k}$.  Note that SSJD can also work with these gain loops, without feeding $\hat{g}^+_{k}$ and $\hat{g}^-_{k}$ to the path metric evaluation. 



\begin{figure}
\centering
\includegraphics[width=0.8\columnwidth]{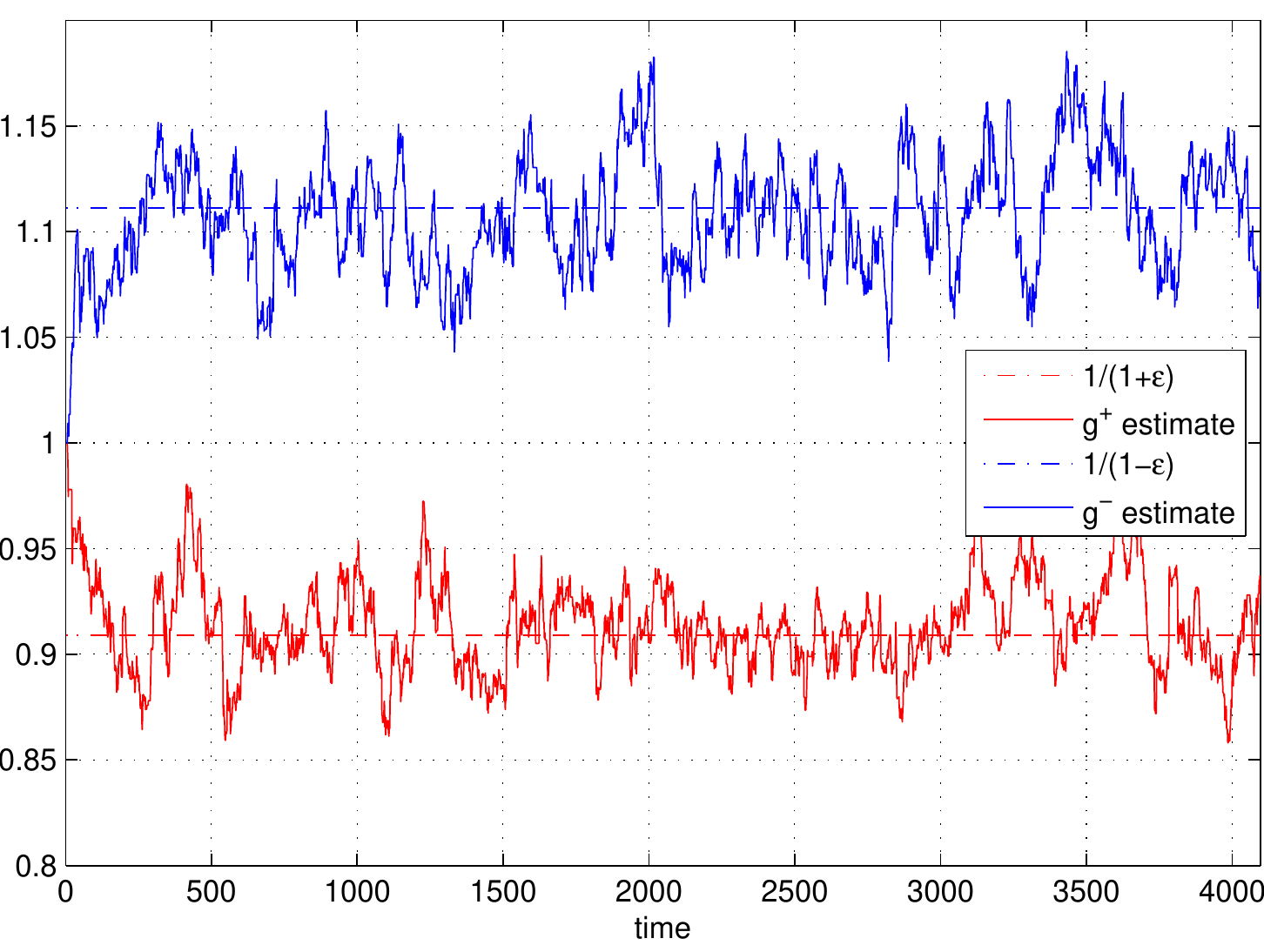}
\caption{Adaptive estimation of $g^+$ and $g^-$ over one sector of 4096 bits on channel $h(d)=1+D$ at $\text{SNR}=10\text{dB}$.}
\label{fig_gtracking}
\end{figure}

\Figref{fig_gtracking} shows the behavior of the $g^+_k$ and $g^-_k$ gain loops in one sector of length $N=4096$ bits on the channel $h(D)=1+D$ at $\text{SNR}=10\text{dB}$ with step-size $\beta=0.005$ and delay $m=5$. For channels with longer memory, a larger delay $m$ may be adopted.

\section{Simulation Results}
\label{sec_simulation}

%
\begin{figure} %
\centering
\subfigure[]{\includegraphics[width = 0.5\columnwidth]{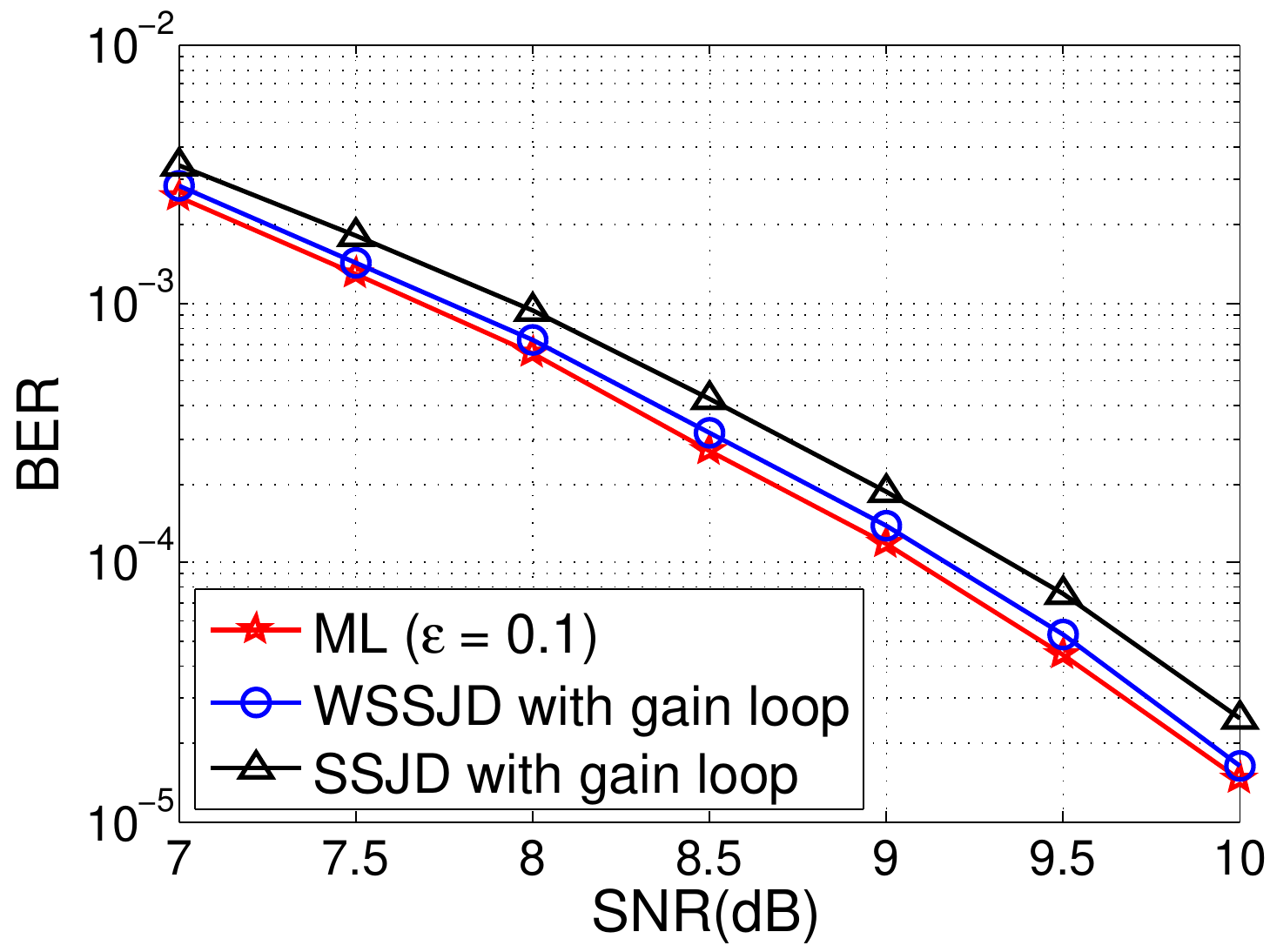}
\label{fig:fixed1}
}~
\subfigure[]{\includegraphics[width = 0.5\columnwidth]{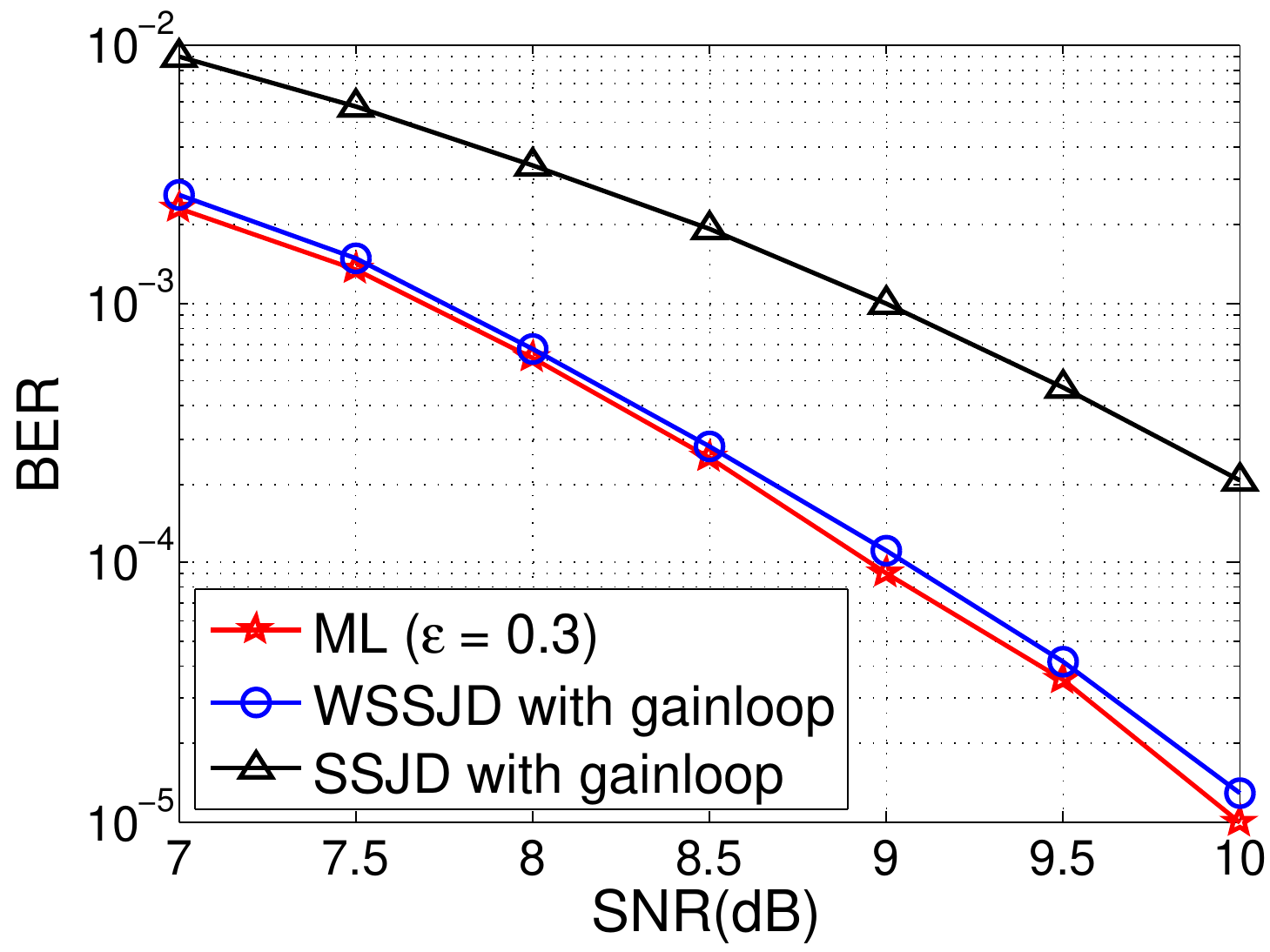}
\label{fig:fixed3}
}

\caption{BER vs. SNR of different detectors with (a) $\epsilon =0.1$ and  (b) $\epsilon=0.3$.} 
\label{fig_fixalpha}
\end{figure}

We simulate WSSJD and SSJD with gain control on the 2H2T system with channel polynomial $h(D)=1+D$. In both cases we set $\beta=0.008$ and $m=5$. The initial values of gain factors $g^+_0$ and $g^-_0$ are obtained by passing training samples through the system. The SNR is defined as
\[
\text{SNR(dB)}=10\log\frac{\|h(D)\|^2}{2\sigma^2 }
\]

We first test the performance of the gain control loops when $\epsilon$ is fixed. \Figref{fig_fixalpha} compares bit error rate (BER) vs. SNR of the ML detector, WSSJD,  and SSJD, for $\epsilon=0.1$ and $\epsilon=0.3$. The frame size is 4096 bits. We assume that the ML detector knows the value $\epsilon$, while WSSJD and SSJD adaptively estimate $\epsilon$ as in \Figref{fig_gainloop}. The static ML detector provides a lower bound for optimal BER performance. It can be seen that adaptive WSSJD performs very close to the static ML detector.  
As expected from the minimum distance plots in \Figref{fig_distance}, the performance of the SSJD is more severely degraded when $\epsilon=0.3$ than when $\epsilon=0.1$. The measures of frame error rate (FER) vs. SNR correlate well with the BER curves in the simulations.


\begin{figure}
\centering                                                     
\subfigure[]{\includegraphics[width = 0.5\columnwidth]{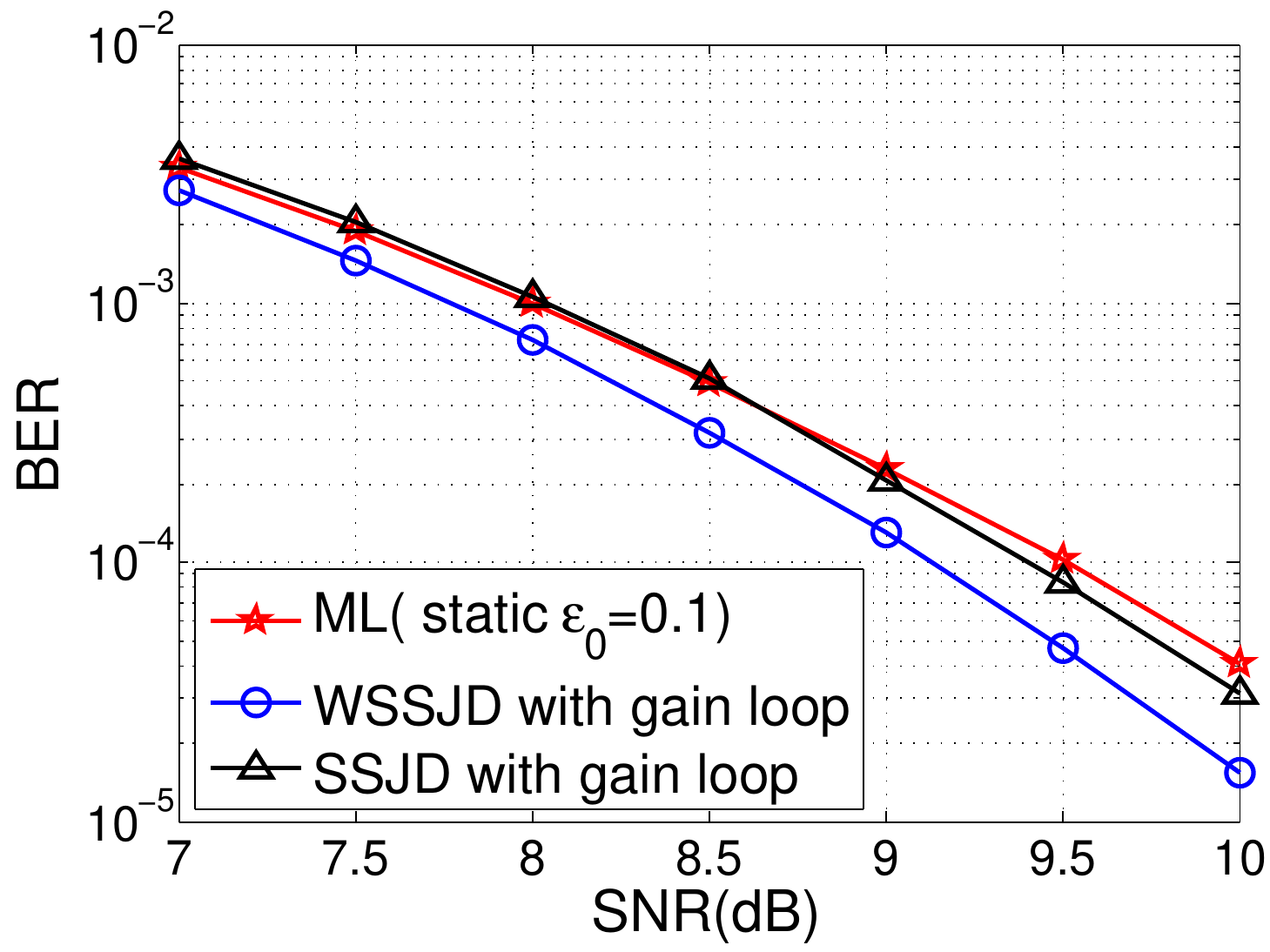}
\label{fig:vary1}
}~
\subfigure[]{\includegraphics[width = 0.5\columnwidth]{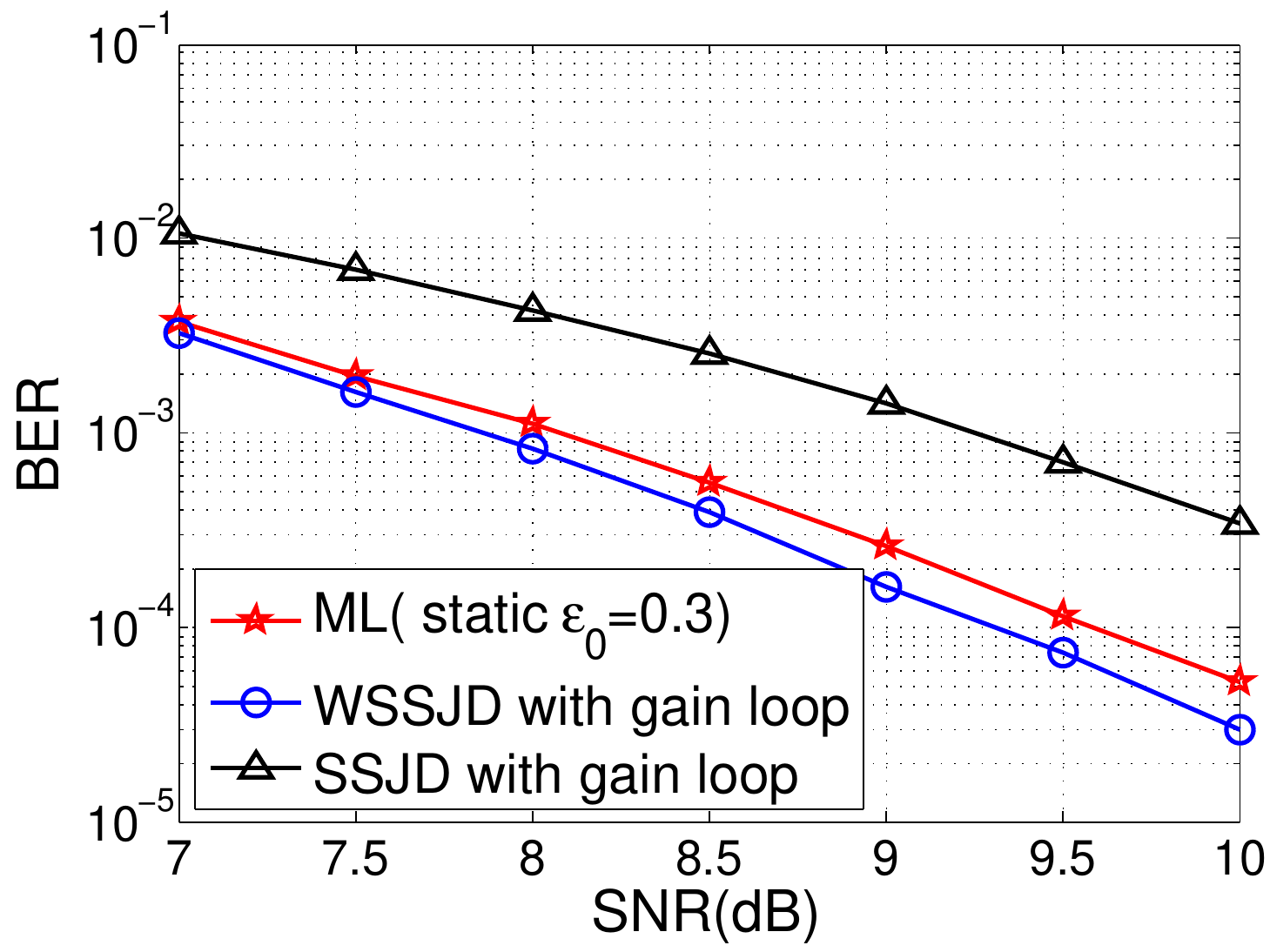}
\label{fig:vary3}
}
\caption{BER vs. SNR of different detectors with $\epsilon$ slowly varying about the mean value (a) $\epsilon_0=0.1$  and (b) $\epsilon_0=0.3$.}
\label{fig_varyalpha}
\end{figure}

Next, we test the performance of the detectors with a dynamic ITI model in which $\epsilon$ changes slowly with respect to the location $k$ in a sector. Specifically, we set
\[
\epsilon(k)=\epsilon_0+0.1\sin(4\pi(k/N))
\]
where $N=4096$ is the frame size and $\epsilon_0$ is the mean ITI value. The ML detector again uses the static value $\epsilon_0$, while WSSJD and SSJD adaptively estimate $\epsilon(k)$. The simulation results, shown in \Figref{fig_varyalpha}, suggest that the adaptive algorithms outperform the static ML detector by about $0.3$-$0.5$dB at high SNR.

In both cases, the performance of a single track detector on $1+D$ channel with no ITI is plotted for comparison. It is interpreted as the best performance an ITI cancellation scheme can achieve, where the detector is assumed to have perfect knowledge about the side track information and the interference parameter $\epsilon$.

\begin{figure}
\centering
\includegraphics[width=0.85\columnwidth]{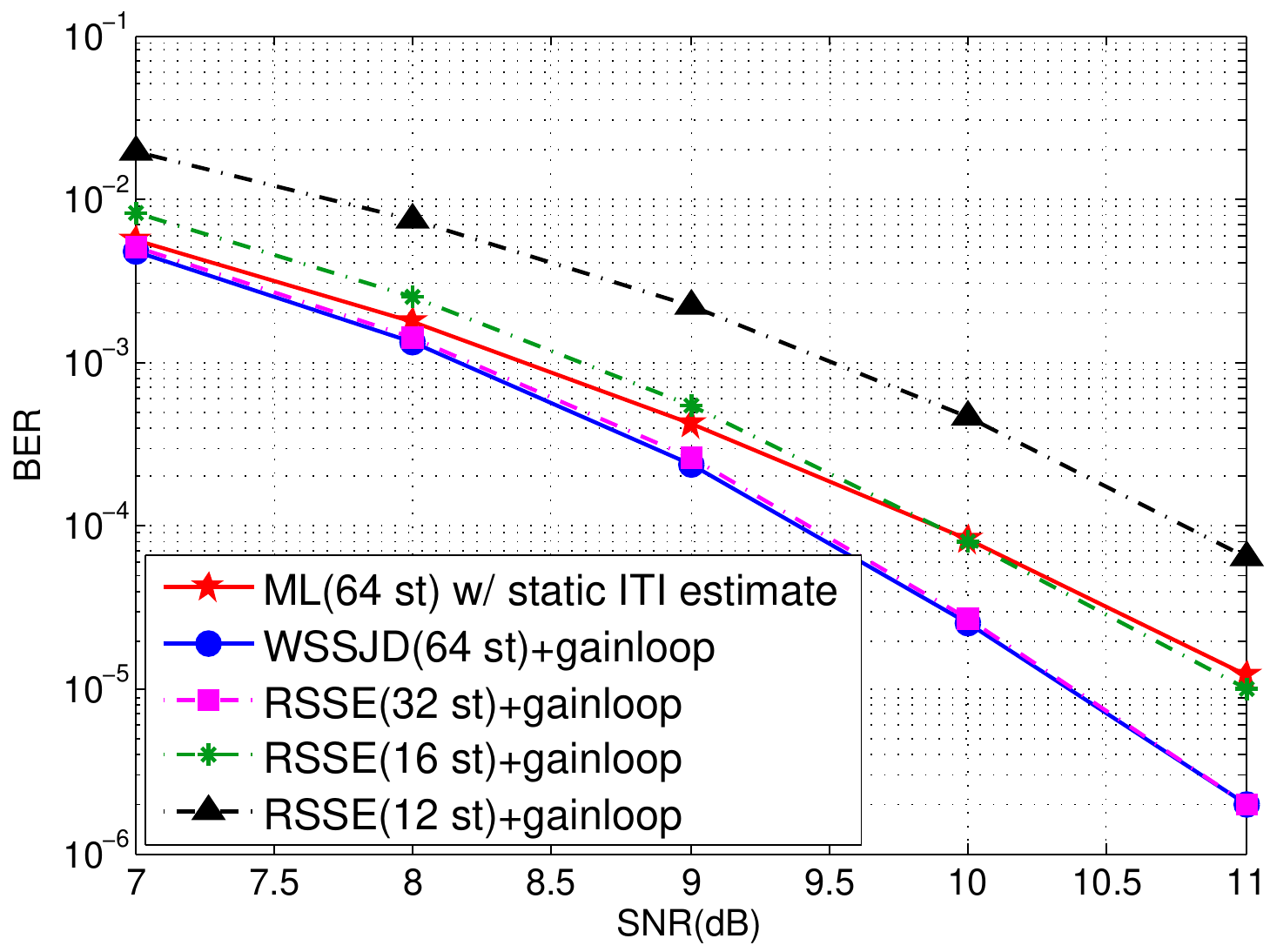}
\caption{Performance of reduced complexity implementations of WSSJD with gain loop on 2H2T EPR4 channel. $\epsilon$ is sinusoidally varying with mean value $\epsilon_0 = 0.1$}
\label{fig_rsse}
\end{figure}

The optimal MHMT detector suffers from high complexity that prevents it to be practical when the channel memory is large. We address this problem In \cite{bing_intermag}. We show that the decomposition method in WSSJD leads to a natural set partition design of the input symbols, based on which the reduced-state sequence estimation (RSSE) algorithm could be applied. Figure \ref{fig_rsse} shows a simulation result for WSSJD with RSSE on EPR4 channel. On this channel, WSSJD outperforms the static ML detector by adapting to the ITI level, and WSSJD+RSSE can achieve nearly identical performance as WSSJD with only 32 states in stead of 64 states.

\section{WSSJD on General ITI Channel}
\label{sec_general}
In next generation magnetic recording disks, the tracks are proposed to be organized in bands \cite{AHLMPS11}. Inside each band, the tracks are squeezed and closely aligned, while between bands there is a small gap to prevent interference. To maintain good performance, a band of tracks will be processed together, with information shared across the tracks. In this section, we generalize the WSSJD algorithm to meet the requirement of detecting $n$-tracks simultaneously. The generalized trellis is independent of the ITI, and gain loops will be used to efficiently get the estimate of ITI. 

\subsection{$n$-Head, $n$-Track Channel}
Consider a band of $n$ tracks. Let $x^i(D)$ denote the bipolar data sequence recorded on the $i$-th track. There are $n$ heads evenly placed over the tracks, reading back simultaneously. An alternative way to read the multiple tracks is to use one head to sequentially scan the track band $n$ times, and each time concentrate on one track. Let $r^i(D)$ denote the sampled read back sequence obtained by concentrating on the $i$-th track for $i = 1, ..., n$. They form an $n$-head $n$-track ($n$H$n$T) system, with 
\begin{align}
\bfX(D)=[x^1(D), \cdots,x^n(D)]^\top
\end{align}
as the input vector, and 
\begin{align}
\bfR(D)=[r^1(D),\cdots,r^n(D)]^\top
\end{align}
as the output vector.
. Assume all the tracks are equalized to the same target $h(D)$. The mathematical relation between $\bfX(D)$ and $\bfR(D)$ is
\begin{align}
\mathbf{R}(D)=A_{n}\mathbf{X}(D)h(D)+\mathbf{\Omega}(D), \label{eq_47}
\end{align}
where $\mathbf{\Omega}(D)=[\omega^1(D), \cdots,\omega^n(D)]^\top$ are the electronic noise components. We assume that the noise samples are independent and Gaussian distributed, with zero mean and variance $\sigma^2$. The term $\bfX(D)h(D) = [x^1(D)h(D), \cdots, x^n(D)h(D)]^\top$ denotes the vector of noiseless ISI channel outputs, and $A_n$ is an $n\times n$ interference matrix. If we only consider the most significant ITI, which comes from the adjacent tracks, and assume the ITIs are symmetric, then $A_n$ can be modeled as a {tridiagonal Toeplitz matrix}
\begin{align*}
A_{n}=\left[\begin{array}{ccccc}
1 & \epsilon &  &  & \\
\epsilon & 1 & \ddots & \text{\Large 0} &\\
 &\ddots & \ddots & \ddots & \\
 &  \text{\Large 0} & \ddots & 1 & \epsilon \\
 &   &  & \epsilon & 1
\end{array}\right],
\end{align*}
where $\epsilon\in[0,0.5]$ is the ITI parameter.

Given the received sequences $\mathbf{R}(D)$, the ML detector chooses $\tilde{\mathbf{X}}(D)$ that satisfies
\begin{align}
\tilde{\mathbf{X}}(D) &= \arg \max_{\mathbf{X}(D)} \text{Pr}(\mathbf{R}(D|\mathbf{X}(D)) \notag\\
&= \arg \min_{\mathbf{X}(D)} \|\mathbf{R}(D)-A_n\mathbf{X}(D)h(D)\|^2. \label{eq_48}
\end{align}
The squared norm of a sequence vector, $\|\bfX(D)\|^2$, is calculated by 
$\|\bfX(D)\|^2 = \sum_i \|x^i(D)\|^2 = \sum_{i,j}(x^i_j)^2$. 
The trellis constructed to find $\tilde{\mathbf{X}}(D)$ in (\ref{eq_48}) contains $2^{n\nu}$ states, each of which is associated with $2^n$ edges. The output labels are calculated from the noiseless ISI channel output $A_n\mathbf{X}(D)h(D)$, thus requiring the knowledge of $\epsilon$. 

For an error event
\begin{align}
\bfe(D)=[e^1(D),\cdots,e^n(D)]^\top,
\end{align}
where $e^i(D) = x^i(D)-\tilde{x}^i(D)$ is the error sequence on the $i$-th track, the distance associated with $\bfe(D)$ is calculated by
\begin{align}
d^2(\bfe(D)) &=  \|A_n\mathbf{e}(D)h(D)\|^2 \notag\\
			 &= \sum_{i=1}^n \|y^i(D)\|^2
\end{align}
where
\begin{align}
y^1(D) &= [e^1(D)+\epsilon e^2(D)]h(D) \\
y^n(D) &= [e^n(D)+\epsilon e^{n-1}(D)]h(D) \\
y^i(D) &= [e^i(D)+\epsilon e^{i-1}(D) +\epsilon e^{i+1}(D)]h(D), i\in[2,n-1]
\end{align}
The minimum distance of the channel is obtained by minimizing $d^2(\bfe(D))$ over all possible $\bfe(D)$. In Figure \ref{fig_distance}, we plot the minimum distances of $3$H$3$T, $4$H$4$T and $5$H$5$T found by computer search. In a large region of $\epsilon$, the $n$H$n$T ML detectors have a greater minimum distance property than the ITI-free SHST ML detector.  

\subsection{Decomposition of Interference Matrix}
The conventional ML detector involves $\epsilon$ in its trellis construction. In this section we will show that by decomposing the channel carefully we can have an ML-equivalent algorithm whose trellis is independent of the ITI level. 

Consider the eigendecomposition of $A_n$, 
\begin{align}
A_{n}=V_n\Lambda_{n}V_n^{\top}, 
\end{align}
where $V_n$ is an $n\times n$ matrix whose columns are the eigenvectors of $A_n$, and $\Lambda_n$ is a diagonal matrix whose diagonal elements are the corresponding eigenvalues.
The eigenvalues and eigenvectors of the symmetric tridiagonal Toeplitz matrix have a known closed form \cite{Smith85}\cite{NPR2012toeplitz}. If we define
\begin{align}
 \hat{T}_{n}=\left[\begin{array}{cccc}
0 & 1 &  & \mathcal{O}\\
1 & 0 & \ddots\\
 & \ddots & \ddots & 1\\
\mathcal{O} &  & 1 & 0
\end{array}\right],
\end{align}
then 
\begin{align}
A_n = I_n + \epsilon \hat{T_n} = V_n(I_n+\epsilon \hat{\Lambda}_n)V_n^\top,
\end{align}
where $I_n$ is an $n\times n$ identity matrix, and $\hat{\Lambda}_n$ is the diagonal matrix containing the eigenvalues of $\hat{T}_n$. Therefore, the columns of $V_n$ are also the eigenvectors of $\hat{T}_n$, and $\Lambda_n=I_n+\epsilon\hat{\Lambda}_n$. In fact, $\hat{\Lambda}$ and $V_n$ have closed forms: the $k^{\text{th}}$ eigenvalue of $\hat{T}_n$ is 
\begin{align}
\hat{\lambda}_k = 2\cos \left(\frac{k\pi}{n+1} \right),
\end{align}
and the $j^{\text{th}}$ element in the $k^{\text{th}}$ eigenvector $\bfv_k$ is 
\begin{align}
v_{jk}=\sqrt{\frac{2}{n+1}}\sin \left(\frac{kj\pi}{n+1}\right).
\end{align}
Note that $V_n$ is independent of $\epsilon$.
\begin{example}
\label{ex_1}
For the case $n=2$,
\begin{align*}
\begin{array}{cc}
\Lambda_{2}=\left[\begin{array}{cc}
1+\epsilon & 0\\
0 & 1-\epsilon
\end{array}\right], & \quad V_{2}=\left[\begin{array}{cc}
\frac{\sqrt{2}}{2} & \frac{\sqrt{2}}{2}\\
\frac{\sqrt{2}}{2} & -\frac{\sqrt{2}}{2}
\end{array}\right]\end{array}
\end{align*}
\end{example}
\begin{example}
\label{ex_2}
For the case $n=3$,
\begin{align*}
\Lambda_{3} &=\left[\begin{array}{ccc}
1+\sqrt{2}\epsilon & 0 & 0\\
0 & 1 & 0\\
0 & 0 & 1-\sqrt{2}\epsilon
\end{array}\right], \\
 V_{3} &=\left[\begin{array}{ccc}
\frac{1}{2} & \frac{\sqrt{2}}{2} & \frac{1}{2}\\
\frac{\sqrt{2}}{2} & 0 & -\frac{\sqrt{2}}{2}\\
\frac{1}{2} & -\frac{\sqrt{2}}{2} & \frac{1}{2}
\end{array}\right].
\end{align*}
\end{example}

\subsection{Channel Decomposition and Generalized WSSJD}
Consider the channel model (\ref{eq_47}). Substituting $A_n$ by its eigendecomposition gives
\begin{align}
\mathbf{R}(D)= V_n\Lambda_{n}V_n^{\top}\,\mathbf{X}(D)h(D)+\mathbf{\Omega}(D). \label{eq_52}
\end{align}
Reorganize (\ref{eq_52}) to get
\begin{align}
\Lambda_{n}^{-1}V_n^{\top}\mathbf{R}(D)=V_n^{\top}\mathbf{X}(D)\,h(D)+\Lambda_{n}^{-1}V_n^{\top}\mathbf{\Omega}(D).
\end{align}
Let $\bar{\bfX}(D)=V_n^{\top}\mathbf{X}(D)$, $\bar{\bfR}(D)=\Lambda_n^{-1}V_n^{\top}\mathbf{R}(D)$ and $\bar{\mathbf{\Omega}}(D)=\Lambda_n^{-1}V_n^{\top}\mathbf{\Omega}(D)$ be the vectors of new input sequences, received sequences and noises, respectively. This transformed channel model becomes
\begin{align}
\bar{\bfR}(D) = \bar{\bfX}(D)h(D)+\bar{\mathbf{\Omega}}(D), \label{eq_60}
\end{align}
which is composed of $n$ parallel channels. The $j$-th channel is obtained by considering the $j$-th row of both sides of equation (\ref{eq_60}), which gives
\begin{align}
\bar{r}^j(D)=\bar{x}^j(D)h(D)+\bar{\omega}^j(D), \label{eq_61}
\end{align}
where
\begin{align}
\bar{r}^j(D) &=\frac{1}{1+\epsilon\hat{\lambda}_{j}}\sum_{i=1}^{n}v_{ij}\, r^{i}(D), \\
\bar{x}^j(D) &= \sum_{i=1}^{n}v_{ij}\, x^{i}(D), \\
\bar{w}^j(D) &= \frac{1}{1+\epsilon\hat{\lambda}_{i}}\sum_{i=1}^{n}v_{ij}\, \omega^{i}(D). \label{eq_64}
\end{align} 
Several properties of these new channels can be observed:
\begin{enumerate}
\item The noise components in $\bar{\mathbf{\Omega}}(D)$ are still independent. Let $\mathbf{\Omega}_i$ and $\bar{\mathbf{\Omega}}_i$ denote the vectors of the original and transformed noise samples at time $i$, i.e., the coefficients of $D^i$ in the sequences $\mathbf{\Omega}(D)$ and  $\bar{\mathbf{\Omega}}(D)$, respectively. Then 
\begin{align}
E[\bar{\mathbf{\Omega}}_i \bar{\mathbf{\Omega}}_i^\top] =E[\Lambda_n^{-1}V_n^{\top}\mathbf{\Omega}\mathbf{\Omega}^{\top}V_n\Lambda_n^{-1}]=\sigma^2(\Lambda_n^{-1})^2,
\end{align}
which is a diagonal matrix. So the components of $\bar{\mathbf{\Omega}}_i$ are uncorrelated and Gaussian, therefore independent. Furthermore, the noise power of the $j$-th channel is $\sigma^2/\lambda_j^{2}$.
\item After the transformation, the inputs of different channels have different alphabets. For the $j$-th channel, the alphabet $\Sigma_j$ is
\begin{align}
\Sigma_j = \{\sum_{i=1}^{n} v_{ij}x_i| x_i\in \{+1, -1\}\}
\end{align} 
\item The $j$-th channel corresponds to transmitting $\bar{x}^j(D)$ through the ISI channel $h(D)$ and adding electronic noise of power $\sigma^2/\lambda_j^{2}$. Since the inputs to different channels are correlated, a joint trellis is needed to search for the optimal decision. The new trellis state can be found by applying the one-to-one mapping $V_n^\top \bfx$ to the conventional ML state $\bfx$. The resulted WSSJD trellis has $2^{n\nu}$ states.
\item Since $V_n$ is determined once $n$ is given, the WSSJD trellis is well-defined, and the branch labels are also independent of $\epsilon$.  
\end{enumerate}

The optimal decision $\bar{\mathbf{X}}^*(D)$ satisfies
\begin{align}
\bar{\mathbf{X}}^*(D)& = \arg\max_{\bar{\mathbf{X}}(D)} \, \log \text{Pr}(\bar{\mathbf{R}}(D)| \bar{\mathbf{X}}(D)) \notag \\
& = \arg\min_{\bar{\mathbf{X}}(D)}\, \sum_{j=1}^{n}\lambda_j^2 \, \|\bar{r}^j(D)-\bar{x}^j(D)h(D)\|^2.
\end{align}
For a given error event $\bar{\mathbf{e}}(D) = [\bar{e}^1(D), ..., \bar{e^n}(D)]$, where $\bar{e}^j(D) = \bar{x}^j(D) - \bar{x}^{j*}(D)$, its distance is calculated by
\begin{align}
d^2(\bar{\mathbf{e}}(D)) = \sum_j \, \lambda_j^2\,\|\bar{e}^j(D)h(D)\|^2
\end{align} 
From the above analysis, it is easy to see that WSSJD gives the optimal ML solution.

\subsection{Gain loops}
As shown in equations (\ref{eq_61})-(\ref{eq_64}), for each channel $\epsilon$ appears in a gain factor normalizing $\sum_{i=1}^{n}v_{ij}\, r^{i}(D)$ such that its expectation is $\bar{x}(D)h(D)$. Gain loops can be used to adaptively estimate these gain factors. 

Let $g^j_k$ denote the gain factor estimated for the $j^{\text{th}}$ channel at time $k$. Then $E[g^j_k] = \frac{1}{1+\epsilon\hat{\lambda}_j}$. The LMS adaptive algorithm for updating $g^j_k$ is
\begin{align}
&\hat{r}^j_k = g^j_{k-1}\sum_{i} v_{ij} r^i_k, \\
& \hat{e}^j_{k-\delta} =  \hat{y}^j_{k-\delta} - \hat{r}^j_{k-\delta}, \\
& g^j_k  = g_{k-1}^j + \beta \hat{y}^j_{k-\delta}  \hat{e}^j_{k-\delta},
\end{align}
where $\hat{y}^j_{k-\delta}$ is the instantaneous decision on the noiseless output of the $j^{\text{th}}$ ISI channel at time $k-\delta$. To find it, pick the trellis state which currently has the smallest path metric, and trace back the path history for $\delta$ time slots to obtain the corresponding channel output. The gain factors $g^j_k$ are also used in weighting the path metric. 

\begin{algorithm}[!t]
\caption{WSSJD with gain loop on $n$H$n$T}
\begin{algorithmic}[1]
\State \textbf{function} $\hat{\bfX}(D)=$WSSJD$(\bfR(D),\epsilon_0)$
\State \textit{Initialize:}
\State $M(0)=0$, \\
$M(p)=\infty$ for $p=1,\cdots, 2^{n\nu}-1$ \Comment{path metric}
\State $\Psi$ = $2^{n\nu}\times L$ zero matrix \Comment{path history}
\State 
$G = (I_n+\epsilon_0\hat{\Lambda}_n)^{-1}$\Comment{gain factors}
\State \textit{Begin:}
\For{$k=1 $ to $L$}
\State $\bar{\bfR}_k = G V_n^{\top} \bfR_k$
\For{$p=0$ to $2^{n\nu}-1$}
\State for each predecessor state $q_i$ of $p$
\State $m_i = M(q_i) + (\bar{\bfR}_k-\bfy^{(q_i, p)})^\top G^{-2} (\bar{\bfR}_k-\bfy^{(q_i, p)})$
\State update $M(p)=\min\limits_i m_i$
\State $\Psi(p, k) = q_i$ \Comment{extend survivor path}
\EndFor
\If{$k>\delta$}
\State $p^*=\arg \min\limits_{p} M(p)$
\For {$j=1$ to $\delta$}
\State $p^* = \Psi(p^*, k-j+1)$;
\EndFor\Comment{trace back path history}
\State $\bfe = \bfy^{(\Psi(p^*, k-\delta), p^*)} - \bar{\bfR}_{k-\delta} $
\State $G = G + \beta \text{diag}(\bfy^{(\Psi(p^*, k-\delta), p^*)}) \text{diag}(\bfe)$
\EndIf
\EndFor
\State \textit{end}
\end{algorithmic}
\label{alg_1}
\end{algorithm}

Algorithm \ref{alg_1} summarizes the procedures to implement WSSJD with gain loop on the $n$H$n$T channel. To improve the readability, some terms are explained here.
\begin{enumerate}
\item $G$ is a diagonal matrix with $g^j_k$ as the diagonal elements.
\item $\bfR_k$ is a column vector of the received signals from $nHnT$ channel at time $k$. $\bar{\bfR}_k$ is the vector of outputs from the transformed channel.
\item $M(p)$ is the accumulated path metric at state $p$.
\item $\bfy^{q,p}$ is a column vector of the trellis output label from state $q$ to $p$.
\item $\text{diag}(\bfv)$ transforms the column vector $\bfv$ to a diagonal matrix, with the vector elements aligned on the diagonal.
\end{enumerate}

\begin{figure}
\centering
\includegraphics[width=0.9\columnwidth]{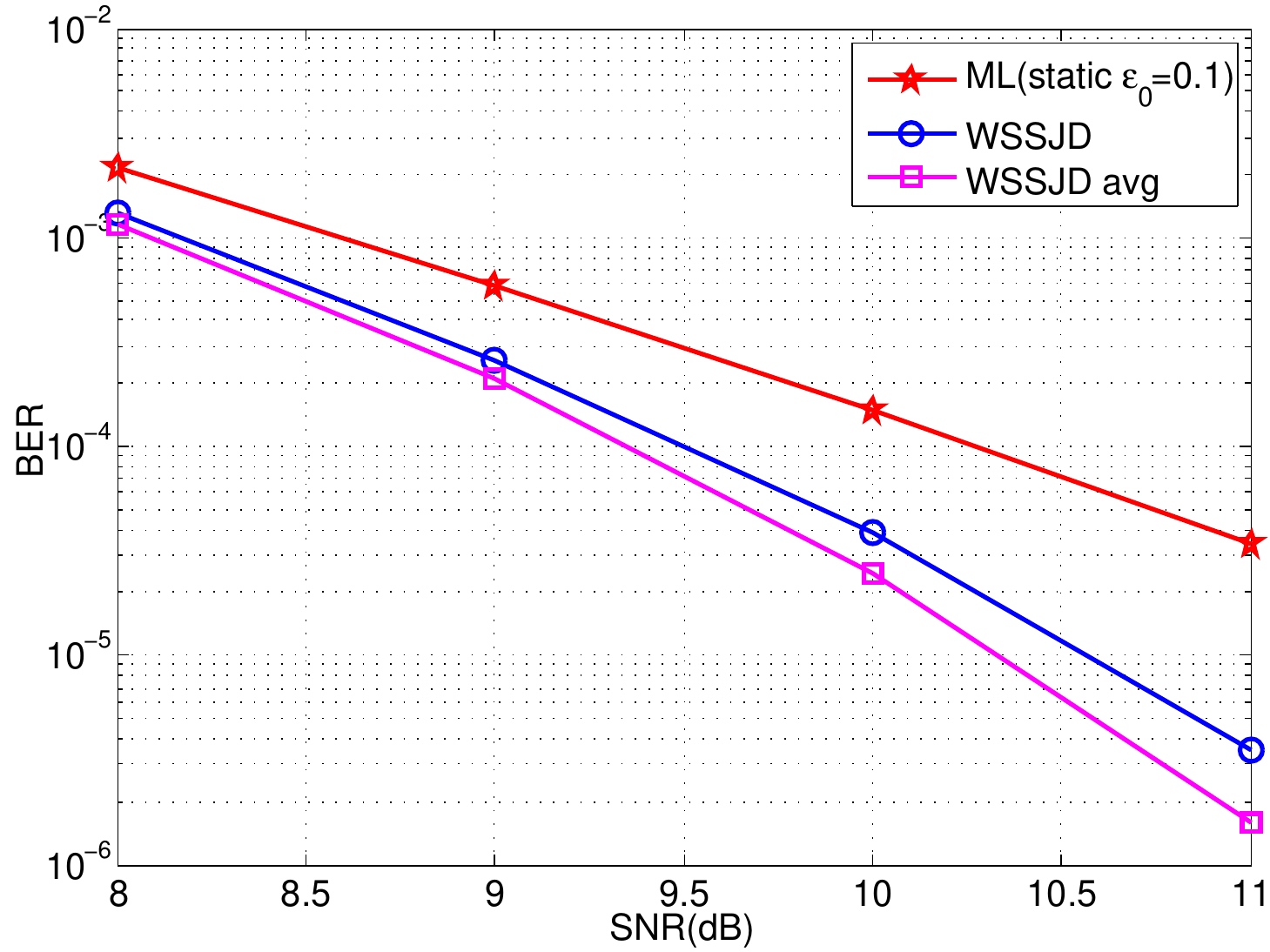}
\caption{BER performance of WSSJD on $3$H$3$T EPR4 channel ($h(D)=1+D-D^2-D^3$)}
\label{fig_3H3T}
\end{figure}

In Figure \ref{fig_3H3T} we plot the BER performance of WSSJD on $3$H$3$T channel, with spatially varying ITI level. Each component channel is equalized to the EPR4 target ($h(D)=1+D-D^2-D^3$). As indicated in Example \ref{ex_2}, the $3$H$3$T channel only requires 2 gain loops, to normalize the first and the third transformed channel. We observe that the WSSJD outperforms the static ML  algorithm by about $1$dB when the BER is in the region of $[10^{-5},10^{-4}]$. The performance can be further improved by averaging the two gain factors, to get a better estimate of $\epsilon$. 
 
\section{Conclusion}
\label{sec_conclusion}
In this paper we propose a novel detector, weighted sum-subtract joint detector (WSSJD), on a generalized $n$H$n$T channel with ITI. The application of channel decomposition transforms the $n$H$n$T channel into $n$ parallel component channels. After the transformation, the ITI level appears as the gain factors on each channel, and can be estimated by gain loops. The proposed algorithm is proved to be ML-equivalent, butcan track small changes in ITI. We specifically investigate the 2H2T case, and analyze the behaviors of several detectors by minimum distance property. The WSSJD technique is also amenable to a reduced complexity implementation. The technique is applicable to the next generation storage systems.

\section*{Acknowledgment}
This work was supported in part by the National Science Foundation under Grant CCF-1405119, and the Center for Memory and Recording Research (formerly, Center for Magnetic Recording Research) at UC San Diego.
\bibliographystyle{IEEEtran}
\bibliography{bibliography_multitrack_phs}

\appendix[Minimum Distance Analysis for ITI Sensitivity]
\label{app}
In this section we give the derivation of equations (\ref{eq_ds}) and (\ref{eq_dd}).

\subsubsection{Single track error events}

Assume $e^b(D)=0$. The distance components reduce to
{\small \begin{align}
& d_{\text{ideal}}=\sqrt{(1+\epsilon_0^2)\|e^a(D)h(D)\|^2} \notag\\
& d_{\text{mism}} = \notag \\ & 2\Delta \epsilon \frac{\left< e^a(D)h(D),x^b(D)h(D)\right> + \epsilon_0 \left< e^a(D)h(D),x^a(D)h(D)\right>}{\sqrt{(1+\epsilon_0^2)\|e^a(D)h(D)\|^2}}.\notag
\end{align}}
We bound $d_{\text{mism}}$ as follows.
\begin{align}
& \left<e^a(D)h(D),x^b(D)h(D)\right> \notag\\
&\quad=\sum\nolimits_n(\sum\nolimits_m x^b_{n-m}h_m)(\sum\nolimits_m e^a_{n-m}h_m) \notag\\
&\quad\leq  |\sum\nolimits_n(\sum\nolimits_m x^b_{n-m}h_m)(\sum\nolimits_m e^a_{n-m}h_m)| \notag\\
&\quad\leq  \sum\nolimits_n|\sum\nolimits_m x^b_{n-m}h_m||\sum\nolimits_m e^a_{n-m}h_m| \notag\\
&\quad\leq  M_h\sum\nolimits_n|\sum\nolimits_m e^a_{n-m}h_m| \notag\\
&\quad=2M_h\sum\nolimits_n|\sum\nolimits_m \frac{e^a_{n-m}}{2}h_m| \notag\\
&\quad\leq 2M_h\sum\nolimits_n (\sum\nolimits_m \frac{e^a_{n-m}}{2}h_m)^2\notag\\
&\quad = \frac{M_h}{2}\|e^a(D)h(D)\|^2 \label{eq_34}
\end{align}
where $M_h=\sum_m|h_m|=2$ for channel $1+D$. Using a similar derivation, we can show
\begin{align}
\left<e^a(D)h(D),x^b(D)h(D)\right> \geq -\frac{M_h}{2}\|e^a(D)h(D)\|^2.
\label{eq_37}
\end{align} 

To find the bounds for $\left<e^a(D)h(D),x^a(D)h(D)\right>$, note that
\begin{align}
&\left< e^a(D)h(D),x^a(D)h(D)\right> \notag\\
&\quad=\sum\nolimits_k (e^a_{k-1}+e^a_{k})(x^a_{k-1}+x^a_{k}) \notag\\
&\quad=\sum_{k=k_1}^{k_2+1} (e^a_{k-1}x^a_{k-1}+e^a_{k-1}x^a_{k}+e^a_{k}x^a_{k-1}+e^a_{k}x^a_{k}) \label{eq_53} \\
&\quad \geq\sum_{k=k_1}^{k_2+1} (|e^a_{k-1}|-|e^a_{k-1}|-|e^a_{k}|+|e^a_{k}|) \label{eq_54} \\
& \quad= 0 \notag
\end{align}
The inequality in (\ref{eq_54}) follows the fact that $x^a_k$ always has the same sign as $e^a_k$, so $e^a_k x^a_k = |e^a_k|$. Choosing $x^a_k$ to have the opposite sign to $e^a_{k-1}$ leads to the lower bound $e^a_{k-1}x^a_{k} \geq -|e^a_{k-1}|$.

The upper bound derived in equation (\ref{eq_34}) is also applicable to $\left< e^a(D)h(D),x^a(D)h(D)\right>$. Therefore,
\begin{align}
0\leq \left< e^a(D)h(D),x^a(D)h(D)\right> \leq \|e^a(D)h(D)\|^2. \label{eq_38}
\end{align}
Combining (\ref{eq_34}) and (\ref{eq_38}), and using $\|e^a(D)h(D)\|\geq 8$ for channel $1+D$, we find that in the case of $\Delta\epsilon>0$ and $\Delta\epsilon<0$:
\begin{align*}
d_{\text{s}} &=d_{\text{ideal}}+d_{\text{mism}} \\
& \geq \left(\sqrt{1+\epsilon_0^2} - \frac{2 \Delta \epsilon}{\sqrt{1+\epsilon_0^2}}\right) \|e^a(D)h(D)\|  \\
& \geq \frac{2\sqrt{2}(1+\epsilon_0^2-2\Delta\epsilon)}{\sqrt{1+\epsilon_0^2}},  \quad \text{if } \Delta\epsilon>0,
\end{align*}
and
\begin{align*}
d_{\text{s}} &=d_{\text{ideal}}+d_{\text{mism}} \\
& \geq \left(\sqrt{1+\epsilon_0^2} + \frac{2\Delta \epsilon(1+\epsilon_0)}{\sqrt{1+\epsilon_0^2}}\right)\|e^a(D)h(D)\| \\
& \geq \frac{2\sqrt{2}(1+\epsilon_0^2+2(1+\epsilon_0)\Delta\epsilon)}{\sqrt{1+\epsilon_0^2}},\quad \text{if } \Delta\epsilon<0.
\end{align*}

These lower bounds are achievable. An example is given in Table \ref{table_2}.

\subsubsection{Double track error events}
In this case, both $e^a(D)$ and $e^b(D)$ are non-zero at some locations. To find an achievable bound on $d_{\text{ideal}}+d_{\text{mism}}$, we assume $\Delta \epsilon\ll 1$. Therefore the distance increment/decrement caused by the mismatch will not be as significant as the distance in the ideal case. The minimum value of $d_{\text{ideal}}$ given by equation (\ref{eq_31}) is $4(1-\epsilon_0)$, achieved by the error sequences of the form
\begin{align*}
\bfe^a &= [0,\cdots, 0, e_{k_1}^a, \cdots, e_{k_2}^a, \cdots, 0]\\
\bfe^b &= [0,\cdots, 0, e_{k_1}^b, \cdots, e_{k_2}^b, \cdots, 0]
\end{align*}
with $e^a_{k+1} = -e^a_k$ for $k_1\leq k \leq k_2-1$, and $e^b_k = -e^a_k$ for $k_1\leq k \leq k_2$. The assumption on $\Delta\epsilon$ suggests that we focus on these error events. We use $d^*_{\text{mism}}$ to denote the minimum distance parameter attained by this subset of double track error events. 

Since
\begin{align}
& \langle \cA(D),x^b(D)h(D)\rangle \notag\\
& =\sum_{k=k_1}^{k_2+1} [e_k^a + e_{k-1}^a +\epsilon_0(e_k^b+e_{k-1}^b)](x_k^b+x_{k-1}^b) \notag\\
& = (e_{k_1}^a +\epsilon_0 e_{k_1}^b)(x_{k_1}^b+x_{k_1-1}^b) + (e_{k_2}^a +\epsilon_0 e_{k_2}^b)(x_{k_2+1}^b+x_{k_2}^b)\notag \\
& = -|e^a_{k_1}|+\epsilon_0 |e_{k_1}^b| + (e_{k_1}^a + \epsilon_0 e^b_{k_1})x^b_{k_1-1} \notag\\
&\quad\quad -|e^a_{k_2}|+\epsilon_0 |e_{k_2}^b| + (e_{k_2}^a + \epsilon_0 e^b_{k_2})x^b_{k_2-1} \label{eq_59}
\end{align}
Upper and lower bounds for (\ref{eq_59}) can be found by carefully assigning values for $x^b_{k_1-1}$ and $x^b_{k_2-1}$. If $x^b_{k_1-1}$ and $x^b_{k_2-1}$ have the same sign as $e^a_{k_1}$ and $e^a_{k_2}$, respectively, (\ref{eq_59}) achieves the maximum value 0. If $x^b_{k_1-1}$ and $x^b_{k_2-1}$ have the same sign as $e^b_{k_1}$ and $e^b_{k_2}$, respectively, (\ref{eq_59}) achieves the minimum value $8(\epsilon_0-1)$. Similarly, we have
\begin{align}
8(\epsilon_0-1)\leq	\langle \cB(D),x^a(D)h(D)\rangle \leq 0.
\end{align}

We conclude that in the case of $\Delta\epsilon>0$ and $\Delta\epsilon<0$:
 \begin{align*}
 d_{\text{d}} &=d_{\text{ideal}}+d^*_{\text{mism}} \\
 & \geq 4(1-\epsilon_0) + \frac{2\Delta\epsilon}{4(1-\epsilon_0)} \cdot 16(\epsilon_0-1) \\
 & = 4(1-\epsilon_0-2\Delta\epsilon), \quad \text{if } \Delta\epsilon>0,
 \end{align*}
 and
  \begin{align*}
  d_{\text{d}} &=d_{\text{ideal}}+d^*_{\text{mism}} \\
  & \geq 4(1-\epsilon_0), \quad \text{if } \Delta\epsilon<0.
  \end{align*}
 
Notice that these bounds are derived for a subset of double track error events which achieve $\min d_{\text{ideal}}$. An example is given in Table \ref{table_3}.

We compared the values of $d^2_{\text{min}} = \min\{d^2_{\text{s}}, d^2_{\text{d}}\}$ obtained by computer search with those computed using $d^*_{\text{mism}}$ as an approximation to $d_{\text{mism}}$, and they agreed at all points plotted in Fig. \ref{fig_dismiss}. This claims that the simplification in our analysis of double track error events does not affect the $d^2_{\text{min}}$ computation.

\end{document}